\documentclass[man,11pt,apacite,noextraspace,longtable,floatsintext]{apa6} 
\usepackage[utf8x]{inputenc} 
\usepackage{graphicx} 
\usepackage{amsmath,amssymb} 
\usepackage{amscd} 
\usepackage{float} 
\usepackage{geometry}
\usepackage{paralist}
\usepackage{caption,subcaption}
\usepackage{booktabs}
\usepackage{wrapfig}
\usepackage{framed}
\usepackage{amsfonts}
\usepackage{bbm}
\usepackage{tikz}
\usepackage{setspace}
\usepackage{rotating,color}

\DeclareDelayedFloatFlavor{sidewaystable}{table}
\usepackage[amsmath,amsthm,thmmarks]{ntheorem} 
\usepackage{geometry}

\theoremstyle{plain}

\title{External Correlates of Adult Digital Problem-Solving Behavior: Log Data Analysis of a Large-Scale Assessment}

\author{Susu Zhang$^1$, Xueying Tang$^2$, Qiwei He$^3$, Jingchen Liu$^4$, and Zhiliang Ying$^4$} 

\affiliation{$^1$University of Illinois at Urbana-Champaign\\
$^2$University of Arizona\\
$^3$Educational Testing Service\\
$^4$Columbia University}

\shorttitle{External Correlates of Digital Problem-Solving Behavior}

\begin{document}

\maketitle

\begin{abstract}

Using the action sequence data (i.e., log data) from the problem solving in technology-rich environments assessment on the 2012 Programme for the International Assessment of Adult Competencies survey, the current study examines the associations between adult digital problem-solving behavior and several demographic and cognitive variables. Action sequence features extracted using multidimensional scaling \cite{tang2019mds} and sequence-to-sequence autoencoders \cite{tang2019seq2seq} were used to predict test-taker external characteristics. Features extracted from action sequences were consistently found to contain more information on demographic and cognitive characteristics than final scores. Partial least squares analyses further revealed systematic associations between behavioral patterns and demographic/cognitive characteristics.

\end{abstract}

\newpage
\section{Introduction}

The increasing popularity of computer-based testing has enabled the administration of interactive test items. As a test-taker attempts an interactive item, their actions are recorded, in the form of a temporally ordered sequence of multi-type, time-stamped events. In contrast to the final response (i.e., a dichotomous or polytomous score), the action sequence preserves the processes that the test-taker goes through to solve the problem. This type of process data can reveal a lot of information about the test-takers, including how they arrived at a final response or the lack thereof, as well as other collateral information, such as their level of engagement.

Various methods for analyzing and making use of process data were recently studied. This includes the examination of how generic process features (e.g., action count and problem-solving time) interact to affect final task completion \cite{voros2016laypersons}, the use of n-gram analysis to extract key actions (or subsequences of actions) associated with the final item responses \cite<e.g.,>{he2018exploring,he2016analyzing}, the use of new measurement models for action occurrences to assess the latent traits of interest \cite<e.g.,>{lamar2018markov, liu2018analysis}, and the modeling of examinee actions as recurrent events \cite<e.g.,>{xu2018latent}, among many others. Most recently, \citeauthor{tang2019mds} \citeyear{tang2019mds} and \citeauthor{tang2019seq2seq} \citeyear{tang2019seq2seq} proposed two generic methods for extracting latent features from action sequences, namely multidimensional scaling (MDS) and sequence-to-sequence autoencoders (Seq2seq). Both feature extraction methods automatically extract numerical features from log data and do not require manual feature engineering a priori using domain knowledge. Although both MDS and Seq2seq encode variable-length, categorical action sequences into finite-dimensional, continuous latent features, the two approaches differ in their objectives. Specifically, MDS aims at preserving the pair-wise differences in individuals' action sequences with the extracted latent features, while Seq2seq aims to reconstruct the original sequences to the greatest extent. With both simulation studies and empirical analyses, it was shown that the extracted latent features could accurately predict the final response of the examinees on the same question. Also, compared to final responses, process features achieved better prediction of examinees' performance on other items and their proficiency on various cognitive traits. 

The extracted features from MDS and Seq2seq, which preserve original action sequence information, can be used not only to draw predictions about test-takers' final item responses or cognitive performance but also to explore how various examinee characteristics, such as demographic or cognitive traits, are associated with their problem-solving behaviors. From a design perspective, this can inform test developers on how an interactive item elicits different test-taking behaviors, which is essential to construct validation and evaluation of item fairness. From an instructional perspective, identifying the individual differences in problem-solving processes can benefit educators who want to understand typical effective and ineffective problem-solving strategies of different subgroups. 

The purpose of the current study is to explore the association between test-takers' external demographic/cognitive characteristics and problem-solving behavior on the problem solving in technology-rich environment (PSTRE) assessment in the 2012 Programme for the International Assessment of Adult Competencies (PIAAC) survey. Specifically, we are interested in several questions, namely (1) the relative strength of association between  external cognitive or demographic characteristics and problem-solving processes, as compared to final scores alone, (2) the difference between the two feature extraction methods (i.e., MDS and Seq2seq) in preservation of external trait information, (3) the differences across items on the extent of association with external traits, and (4) in the case that some items' action sequences associate stronger with a trait than other items, the specific item characteristics and behavioral patterns that explain the associations. The current study quantifies the strength of association with an external trait (e.g., age) as the prediction power of the trait using a specific type of problem-solving data (i.e., final score alone, MDS process features, or Seq2seq process features). For each external trait, the strengths of association of different items were further compared, and some of the most informative items (i.e., items with strongest association with a trait) were examined in more details. This is done by extracting principle features that best explain the covariance between the predicted trait and the process features, and seeking empirical interpretations of the principle features.  

The associations between PIAAC background variables and the process data on the PSTRE interactive items have been explored in several previous studies \cite<e.g.,>{liao2019mapping,he2018background}. The approach adopted in the current study differs from prior research in several ways: While previous studies focused on custom defined variables (e.g., time until first action, total time, number of actions, counts for particular key actions) or short subsequences of consecutive actions (e.g., n-gram features) extracted from log data, in the current study, process features were extracted directly from the full sequences via MDS and Seq2seq. On one hand, both MDS and Seq2seq feature extraction methods are fully automated and preserve more information about long-term events or other key patterns not predefined. On the other hand, the extracted MDS and Seq2seq features are not directly tied to empirical interpretations. To interpret these features, partial least squares (PLS) analysis is performed to project the action features onto directions of high covariance with the dependent variable, and the top PLS components are further used to identify patterns in the original sequences.

% Based on the prediction results, we examined whether these PSTRE items, although all designed to measure the same PSTRE construct, exhibited any notable differences on the prediction power for specific traits. Followed by the evaluation of general prediction power of PSTRE items, we further delved into the differences in individual items' prediction power and evaluated why the action sequences from some items were more informative than others. We sought interpretations for the features with high prediction power and tried to identify specific patterns in the action sequences that were associated with specific test-taker characteristics. 

%The PSTRE items require the test-takers to navigate through one or multiple virtual computer environment(s) to solve given tasks. These environments are designed to assimilate commonly seen computer environments in day-to-day life and workplace, such as email boxes, spreadsheets, and web-browsers. Tasks required by the items involved sorting or sending emails, purchasing an item through a website, and filling in or identifying information from spreadsheets, etc. Students' final scores on each item were derived from predetermined scoring rubrics. In addition to the PSTRE items, the test-takers were also assessed on literacy and numeracy, and a variety of background questionnaires were administered to collect basic demographic information and employment and education status. 

The rest of the paper is organized as follows. We first provide a description of the PSTRE data and other external variables used in the current study. The next section provides an introduction to the feature extraction methods based on MDS and Seq2seq. The Method section details the regularized regression method for background and cognitive variable prediction, evaluation criteria, and the PLS approach to feature interpretations. Findings from empirical analyses and selected feature interpretations are presented in the Results section. A discussion of the empirical findings, as well as their practical implications for assessment design and interventions, is provided at the end.

\section{The PIAAC PSTRE and Survey Data}

The PIAAC \cite<e.g.,>{schleicher2008piaac} is an international study carried out by the Organization for Economic Co-operation and Development (OECD) to assess the cognitive and workplace skills of working-age individuals worldwide. In its first cycle in 2012, participants from 24 countries aged between 16 and 65 years were measured on a selection of information-processing skills, including literacy, numeracy, and PSTRE. In addition to the three cognitive assessments, participants were also administered a series of survey items on their demographic backgrounds, occupational and educational information, and their use of different skills (e.g., reading, writing, information technology) at home and at work, et cetera.

\subsection{Process Data from the PSTRE Assessment}

The process data analyzed in the current study came from the computer-based assessment of PSTRE in the 2012 PIAAC study. For each PSTRE item, the test environment resembled commonly seen informational and communicative technology (ICT) platforms, such as e-mail client, web browser, and spreadsheet. Test-takers were prompted to complete specific tasks on these interactive platforms. Under the PIAAC framework, PSTRE is defined as the use of digital technology, communication tools, and internet to obtain and evaluate information, communicate with others, and perform practical tasks \cite{organisation2012literacy}. Thus, successful responses to the PSTRE items would require not only familiarity with ICT tools and platforms but also the ability to obtain, process, and communicate information in digital environments. 

\begin{figure}
    \centering
    \includegraphics[width = \textwidth]{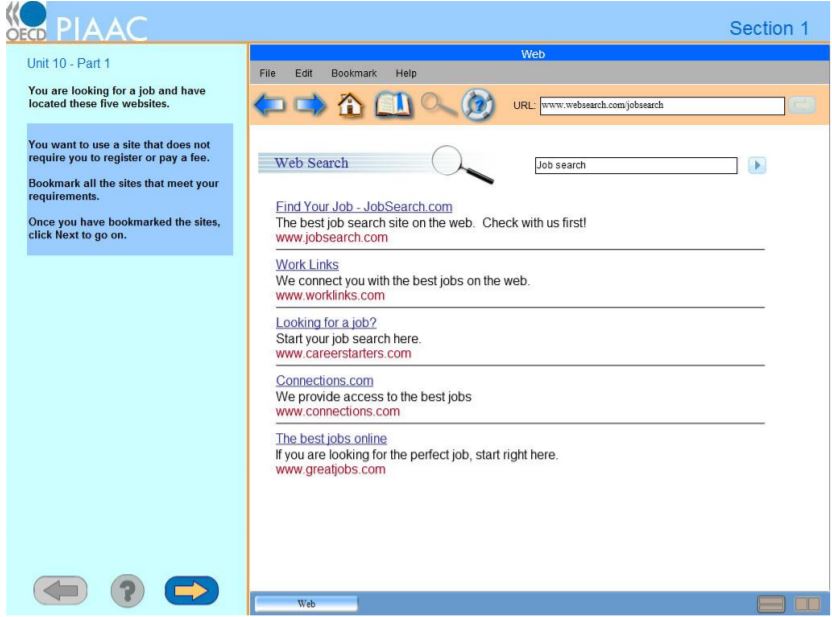}
    \caption{Home page of the PSTRE sample item.} %Reprinted from Sample Questions and Questionnaire, \textit{OECD}. Retrieved February 27, 2019 from   \url{http://www.oecd.org/skills/piaac/Problem\%20Solving\%20in\%20TRE\%20Sample\%20Items.pdf}.
    \label{fig:sample_item1}
\end{figure}

\begin{figure}
    \centering
    \includegraphics[width = \textwidth]{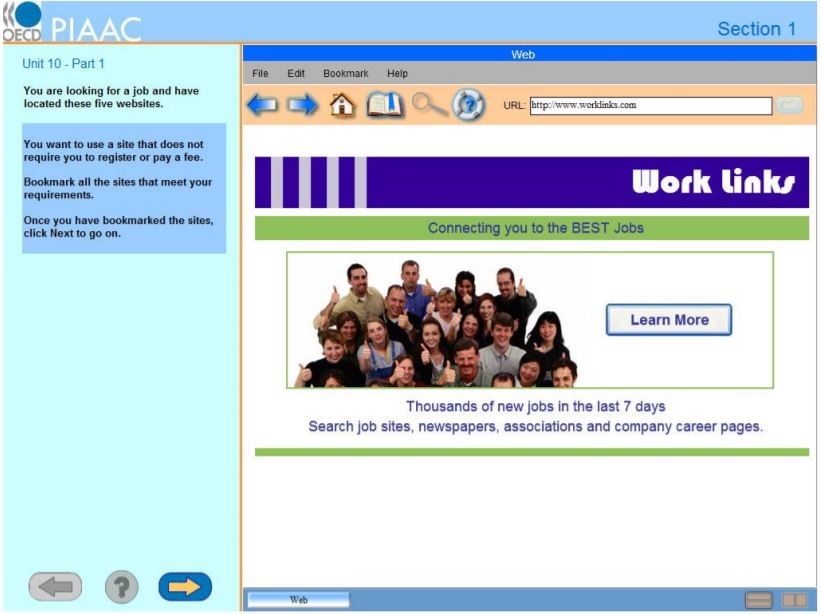}
    \caption{Web page returned from clicking the second link (i.e., ``Work links'') on the home page.}
    \label{fig:sample_item2}
\end{figure}

\begin{figure}
    \centering
    \includegraphics[width = \textwidth]{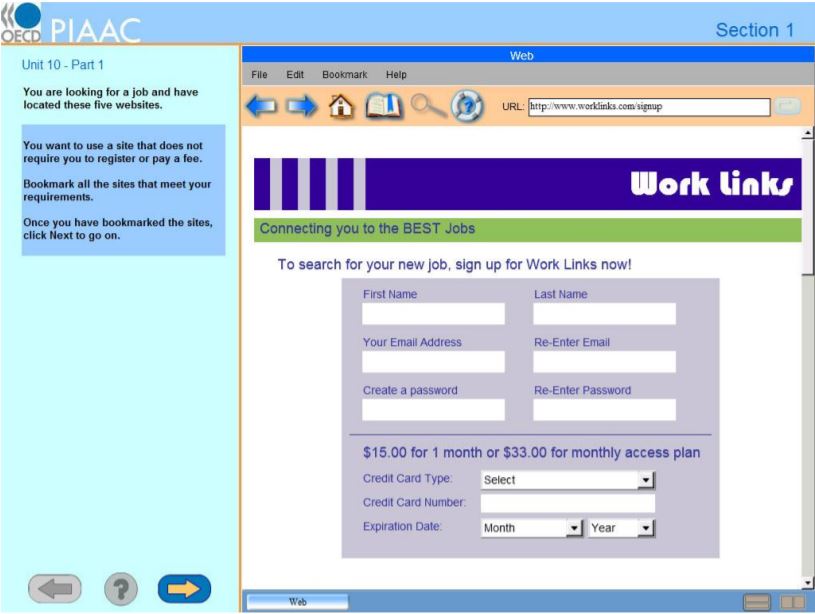}
    \caption{Web page returned from clicking ``Learn More'' on the ``Work links'' website.}
    \label{fig:sample_item3}
\end{figure}

A sample item that resembles PSTRE tasks is shown in Figure \ref{fig:sample_item1}: the instructions for the tasks are presented on the left side of the screen, and on the right is a simulated web browser, with similar functionalities as typical web browsers, such as clicking links, going back and forth, and bookmarking pages. For this particular item, test-takers are presented with five web pages returned from a search of ``Job search'' and are asked to bookmark all pages that do not require registration or fees. By clicking on each link, test-takers will be directed to the corresponding website. For example, by clicking the second link, ``Work Links'', examinees will be directed to Figure \ref{fig:sample_item2}. And by further clicking on the ``Learn More'' button, they will be directed to the webpage shown in Figure \ref{fig:sample_item3}. When examinees have finished,  they can leave the item by clicking on the right arrow icon (``Next'') below the item instructions, where they will be presented with a pop-out window with two options, namely confirming exit (recorded as ``Next\_OK'') or returning to the task (recorded as ``Next\_Cancel'').
%On Figure \ref{fig:sample_item3}, examinees will discover that, for this particular website, searching for jobs will require signing up and purchasing access plans. Based on this information, examinees will learn that this website does not satisfy the requirements for this item. Examinees can return to the initial page by clicking the home icon in the toolbar or return to the last page by clicking the left arrow icon in the toolbar. Alternatively, they can bookmark this page using either the toolbar icon or the drop-down menu, which will lead to an incorrect decision.

The sequence of clicks and keystrokes (if applicable) by the test-taker on the simulated browser are recorded as an action sequence, with ``Start'' at the beginning indicating the start of the question, and ``Next, Next\_OK'' at the end, indicating the examinee has confirmed to move to the next item. For example, if a test-taker clicked the second link (``Work links'') on the initial page, clicked ``Learn More'' on the next page (Figure \ref{fig:sample_item2}), clicked the ``Back'' button on the toolbar twice to return to the home page, and clicked ``Next'' in the left panel and ``OK'' in the pop-up window, this test-taker's action sequence will be recorded as ``Start, Click\_W2, Click\_Learn\_More, Toolbar\_Back, Toolbar\_Back, Next, Next\_OK''.   

In the current study, we analyzed the examinees' action sequences from $14$ PSTRE items. These items shared similar environments as the one exemplified above, and both the polytomous final outcomes and the action sequences of the subjects were recorded. The data from $3645$ examinees, including $1656$ male and $1989$ female examinees, were used. The examinees' age ranged between $16$ and $66$ years, with an average of $38.93$ years, and they came from five countries, including the United Kingdom (England and Northern Ireland), Ireland, Japan, the Netherlands, and the United States of America. These examinees completed all $14$ PSTRE items, presented as two separate blocks of $7$ items each \footnote{In the PIAAC computer-based assessment, examinees were randomly routed to modules of PSTRE, literacy, or numeracy items. The current study used participants who received two PSTRE modules.}. Examinees were additionally administered a background survey that covered their educational status (e.g., total years of education, highest level of education), vocational status (e.g., employment, income, job type), and their use of different skills (e.g., reading, writing, numeracy, and ICT) at home and at work. In the present study, we attend to the participants' age, gender, income, use of ICT skills at home and at work, and years of education. 

Table \ref{tab: PSTRE_items} presents a brief description of the $14$ PSTRE items, including task name, the types of environments involved, percent of individuals who received full credit, minimum action sequence length, maximum sequence length, average sequence length, and the number of action types. Some of the tasks required the examinees to switch between environments. For instance, in item U23: Lamp Return, examinees needed to navigate between the web browser and the email client to obtain necessary information and solve the task. For these questions that involve multiple environments, examinees could switch to a particular environment by clicking the corresponding pane label (e.g., ``Web'', ``Email'') located at the bottom of the test screen. 

\begin{table}[ht]
\caption{Descriptive information of the $14$ PSTRE items.}
\label{tab: PSTRE_items}
\centering
\begin{tabular}{llllrrrr}
  \hline
 Item ID & Item name & Environment & $p$ &$\min(L)$ & $\max(L)$ & $\text{mean}(L)$ & $N$ \\ 
  \hline
 U01a & Party Invitations & Email & 0.56 &   3 & 114 & 18.21 &  40 \\ 
 U01b & Party Invitations & Email & 0.50 &   3 & 132 & 25.65 &  47 \\ 
 U02  & Meeting Room & Email/Web & 0.14 &   3 & 153 & 25.81 &  96 \\ 
 U03a & CD Tally & Web/SS & 0.39 &   3 &  51 & 8.96 &  67 \\ 
 U04a & Class Attendance & Email/SS & 0.12 &   3 & 304 & 37.38 & 636 \\ 
 U06a & Sprained Ankle & Web & 0.27 &   3 &  57 & 10.04 &  30 \\ 
 U06b & Sprained Ankle & Web & 0.53 &   3 &  68 & 15.39 &  26 \\ 
 U07  & Book Order & Web & 0.49 &   3 &  79 & 19.05 &  40 \\ 
 U11b & Locate Email & Email & 0.23 &   3 & 256 & 25.17 & 124 \\ 
 U16  & Reply All & Email & 0.61 &   3 & 267 & 32.98 & 362 \\ 
 U19a & Club Membership & Email/SS & 0.76 &   3 & 357 & 17.64 &  85 \\ 
 U19b & Club Membership & Web/SS & 0.48 &   3 & 396 & 19.38 & 244 \\ 
 U21  & Tickets & Web & 0.38 &   3 &  77 & 19.74 & 138 \\ 
 U23  & Lamp Return & Web/Email & 0.37 &   3 & 138 & 21.62 & 136 \\ 
   \hline
\end{tabular}
\raggedright

\textit{Note:} $p$ refers to the proportion of subjects who answered the item correctly; $\min(L), \max(L),$ and $\text{mean}(L)$ refer to the minimum, maximum, and average sequence length across all subjects, respectively; $N$ refers to the number of types of actions on the item; SS: Spreadsheet. 
\end{table}

\subsection{Background and Cognitive Variables}

The 2012 PIAAC background questionnaire collected participants' responses on hundreds of items, which covered their education, social background, engagement in literacy, numeracy, and ICT use at home and at work, language background, employment information, and some non-economic information, such as health status and political efficacy \cite{irwinforeword}. Six background variables were considered, including the test takers' reported age, gender, logarithm of hourly income (log(Income)), extent of ICT skill use at home (ICTHome), ICT skill use at work (ICTWork), and total years of education (YRSEdu). To adjust for country effects on income, each individual's hourly income was first converted to US dollars based on 2012 exchange rates and then centered based on the observed median in his/her country. Similarly, the country effects in YRSEdu was adjusted by subtracting the country's median from each test-taker's years of education. Scores on ICTHome and ICTWork were derived based on test-takers' responses to a series of corresponding survey items. As some of the tasks appeared to also involve numericy skills (e.g., spreadsheets), performance (i.e., plausible values\footnote{The PIAAC survey derived plausible values of individuals' cognitive performance (e.g., numeracy performance) based both on the responses to cognitive assessments and background covariates. The numeracy performance variable used in the current study is based on the numeracy plausible values recorded in the official PIAAC data.}) on numeracy was also included as an external cognitive variable. We note that missingness exists for some of the background variables, including income, ICTHome, ICTWork, and YRSEdu. For predicting a particular variable, individuals with missing values were case-wise deleted. The mean, standard deviation, and number of non-missing observations of each continuous variable, as well as their pairwise Pearson correlations, are reported in Table \ref{tab:bg_ques_descriptive}. Compared to male participants, female participants on average showed lower log hourly income, ICTWork, and numeracy performance, with small Cohen's $d$ effect sizes. Additionally, the correlations between each variable and the overall PSTRE performance (derived from PSTRE item responses) are also presented. 

\begin{table}[ht]
\centering
\caption{Descriptive statistics of the background variables and cognitive scores.}
\label{tab:bg_ques_descriptive}
\begin{tabular}{rrrrrrrr}
  \hline
   &Age    & log(Income)    & ICTHome    & ICTWork    & YRSEdu    & Numeracy & PSTRE \\ \hline
$N$ &3645   & 2214  &  3375 &  2289&  3479 & 3645 & 3645 \\ 
Mean &  38.930  & -3.581 & 2.076 & 2.081 &  -0.135 & 0.443  &-0.173\\
SD   & 13.546   &  0.404 & 0.937 &1.012&  2.618  & 0.750& 0.748 \\
{\sl Correlation}:      \\
Age & - & 0.269 & -0.111 & 0.065 & 0.018 & -0.05 & -0.307 \\ 
  log(Income) & - & - & -0.047 & 0.31 & 0.318 & 0.276 & 0.149 \\ 
  ICTHome & - & - & - & 0.335 & 0.181 & 0.174 & 0.297 \\ 
  ICTWork & - & - & - & - & 0.22 & 0.176 & 0.245 \\ 
  YRSEdu & - & - & - & - & - & 0.35 & 0.315 \\ 
  Numeracy & - & - & - & - & - & - & 0.796 \\ 
   \hline
\end{tabular}
\textit{Note.} $N$ stands for the number of non-missing observations for each variable.  
\end{table}

\section{Feature Extraction from Action Sequences}

Based on the descriptive statistics of the action sequences presented in Table \ref{tab: PSTRE_items}, it could be observed that great variability existed in the action sequences on each item: The number of actions performed by each individual varied widely, and the size of the pool of possible actions for each item ranged from $26$ to $636$. 
Compared to structured item response or reaction time data, for which numerous models have been proposed by psychometrics researchers, action sequence data pose several additional challenges for analysis: First, an action sequence is a list of discrete events taken by the test-taker. Test-takers differ in their levels of engagement, abilities, and strategies, and as a consequence, the lengths of their actions sequences, as well as the choices of actions they take, can differ widely. In addition, the sequential ordering of the actions are empirically meaningful, as certain subsequences of actions can only be performed in a specific order, and the order in which an examinee performs a series of subtasks may reflect his/her thought process. As discussed in \citeauthor{tang2019mds} \citeyear{tang2019mds}, an effective method for analyzing action sequence data should (1) be able to deal with sequences of categorical events, (2) allow for individual differences in observed sequence lengths, and (3) exploit the information from the sequential ordering of events.

Although several methods for characterizing individuals based on action sequences have emerged in recent years \cite<e.g.,>{he2018exploring, lamar2018markov, chen2019statistical}, the current study only considers two recently proposed methods that encode whole action sequences into finite-dimensional continuous latent traits, namely multidimensional scaling \cite<MDS, >{tang2019mds} and sequence-to-sequence autoencoder \cite<Seq2seq, >{tang2019seq2seq}. In what follows we introduce these feature extraction methods in details. 

\subsection{Multidimensional Scaling}

The purpose of MDS is to find a finite dimensional representation of the observed data that can best preserve the pairwise dissimilarities between individual observations \cite<e.g.,>{borg2003modern}. If we denote the observed data of individual $i\in \{1,\ldots, N\}$ by $s_i$, the typical aim of MDS is to find a mapping of each $s_i$ to a $K$-dimensional numerical vector $\mathbf{x}_i\in \mathbb{R}^K$ which minimizes
\begin{equation}\label{mds_obj}
    \sum_{i = 1}^N\sum_{j = i+1}^N (d_{ij} - \|\mathbf{x}_i - \mathbf{x}_j\|)^2,
\end{equation}
where $d_{ij} = d(s_i,s_j)$ is the dissimilarity between observations $s_i$ and $s_j$ according to some predefined distance measure $d(\cdot)$, and $\|\mathbf{x}_i - \mathbf{x}_j\| = \sqrt{(\mathbf{x}_i - \mathbf{x}_j)^\prime (\mathbf{x}_i - \mathbf{x}_j)}$ is the Euclidean distance between $\mathbf{x}_i$ and $\mathbf{x}_j$. In the case of process feature extraction, $s_i$, $s_j$ are the observed action sequences of individuals $i$ and $j$. Intuitively, MDS transforms each observation into a point in the $K$-dimensional Euclidean space, so that observations that are similar to each other (i.e., low on $d_{ij}$) are close as well in the Euclidean space (i.e., low on $\|\mathbf{x}_i - \mathbf{x}_j\|$), and observations that are dissimilar are farther apart in the Euclidean space after the transformation. MDS has been previously applied to psychometrics research problems to understand individual differences on various cognitive and non-cognitive tasks \cite<see >{takane200611}, but MDS-based analysis of action sequences in psychometric assessments was not studied until recently \cite{tang2019mds}.

Essential to MDS is the choice of an appropriate distance measure, $d$, to capture the dissimilarity between individual observations. When observations are action sequences on interactive items, \citeauthor{tang2019mds} \citeyear{tang2019mds} proposed to use an order-based sequence similarity measure \cite<OSS;>{gomez2008similarity}, which allows for the quantification of the dissimilarity between ordered, categorical, variable-length event sequences. Denote the observed action sequence of the $i$th test-taker by $\mathbf{s}_i = (s_{i1},\ldots, s_{iL_i})$, where $s_{it}$ is the $t$th action performed by the subject, $L_i$ is the total number of actions, and $L_i^a$ is the number of occurrences of a particular action $a$. The dissimilarity between the action sequences of any two test-takers, $d_{ij}$, is given by 
\begin{eqnarray*}
    d_{ij}&=& \frac{f(\mathbf{s}_i,\mathbf{s}_j)+g(\mathbf{s}_i, \mathbf{s}_j)}{L_i+L_j}, \text{ where} \\
    f(\mathbf{s}_i,\mathbf{s}_j) &=& \frac{\sum_{a\in C_{ij}}\sum_{m=1}^{K_{ij}^a}|s_i^a(m)-s_j^a(m)|}{\max\{L_i, L_j\}}, \text{ and }\\
        g(\mathbf{s}_i,\mathbf{s}_j) &=& \sum_{a\in U_{ij}}L_i^a + \sum_{a\in U_{ji}}L_j^a.
\end{eqnarray*}
Here, $C_{ij}$ is the set of common actions occurred in both sequences, $K_{ij}^a = \min\{L_i^a, L_j^a\}$, $s_i^a(m)$ is the serial position of the $m$th occurrence of event $a$, and $U_{ij}$ is the set of unique actions taken by test-taker $i$ but not by $j$. Intuitively, $f(\mathbf{s}_i, \mathbf{s}_j)$ quantifies how the serial positions of common actions in the two sequences differ, and $g(\mathbf{s}_i, \mathbf{s}_j)$ counts the number of actions unique to each sequence. In turn, the dissimilarity between $i$ and $j$, $d_{ij}$, takes into account the differences both in the ordering of same actions and the types of actions taken.

After calculating the dissimilarities ($d_{ij}$) between each pair of individuals' action sequences, the optimization problem that minimizes Equation \eqref{mds_obj} is solved to find $\mathbf{x}_{1},\ldots,\mathbf{x}_{N}$. As the transformed $\mathbf{x}$'s preserve as much as possible the pairwise dissimilarities in the observed action sequences, $\mathbf{x}_i$ could be seen as a $K-$dimensional latent feature vector which contains information of the original action sequence $\mathbf{s}_i$. Readers are referred to \citeauthor{tang2019mds} \citeyear{tang2019mds} for more details on the OSS distance, an illustrative example on how the distance is calculated, and the algorithms for performing the MDS and choosing the dimension of $\mathbf{x}$, $K$.

\subsection{Sequence-to-sequence autoencoder}

Another approach to feature extraction from variable-length action sequences, which implements a sequence-to-sequence autoencoder (Seq2seq), was proposed in \citeauthor{tang2019seq2seq} \citeyear{tang2019seq2seq}. Commonly used for information compression from  phrases or sentences in natural languages, an autoencoder seeks to encode categorical event sequences into lower-dimensional latent vectors, which can then be used to restore the original sequences. A Seq2seq autoencoder is an artificial neural network with two main components, an encoder function $\phi(\cdot)$ that transforms the original input sequence $\mathbf{s}_i$ into a fixed-dimensional latent vector $\boldsymbol{\theta}_i$, and a decoder function $\psi(\cdot)$ that maps the latent vector $\boldsymbol{\theta}_i$ to a reconstructed version of the original sequence, $\hat{\mathbf{s}}_i$. The action sequence feature extraction procedures proposed in \citeauthor{tang2019seq2seq} \citeyear{tang2019seq2seq} employed a recurrent neural network-based autoencoder with multiple hidden layers. The model is structured as follows.

In the encoding stage, each action in the sequence is mapped to a $K-$dimensional continuous representation (i.e., an embedding, $\mathbf{e}_{it}$) based on the surrounding context using an embedding layer \cite{mikolov2013efficient}. 
%For each possible action $a\in \mathcal{A}$, based on the surrounding context of the action $a$ in the observed sequences, the embedding layer finds a continuous representation of the action in a $D-$dimensional Euclidean space, $\mathbf{e}_a \in \mathbb{R}^D$. In this way, each discrete action is now represented by a continuous vector, which allows for subsequent numerical operations. For a test-taker with observed action sequence $\mathbf{s}_i = (s_{i1}, \ldots, s_{iL_i})$, his/her sequence is mapped to an equal-length sequence of $D-$dimensional vectors, $(\mathbf{e}_{i1}, \ldots,\mathbf{e}_{iL_i})$, where $\mathbf{e}_{it} = \mathbf{e}_a$ if $s_{it} = a$. 
Following the action embedding, a recurrent layer is applied to the embedded sequences. The recurrent layer creates a $K-$dimensional hidden state ($\boldsymbol\theta_{it}$) for each time step $t = 1,\ldots, L_i$. The hidden state at time $t$, $\boldsymbol\theta_{it}$ is a function of the hidden state at the previous time step, $\boldsymbol\theta_{it-1}$, and the embedded action performed at time $t$, $\mathbf{e}_{it}$, that is, $
    \boldsymbol\theta_{it} = f(\boldsymbol\theta_{it-1}, \mathbf{e}_{it}).$
Different choices for the recurrent function, $f(\cdot)$, have been proposed, including the long short-term memory \cite<LSTM; >{hochreiter1997long} architecture and the gated recurrent unit \cite<GRU;>{cho2014properties} architecture. For feature extraction from action sequences, \citeauthor{tang2019seq2seq} \citeyear{tang2019seq2seq} adopted the GRU for the recurrent function, which learns from the observed data to update or reset the hidden states depending on the context and the action type. In this way, the recurrent states ($\boldsymbol\theta_{i1}, \ldots, \boldsymbol\theta_{iL_i}$) accumulate information in the action sequence over time, and the hidden state at the final time step, $\boldsymbol\theta_{iL_i}$, summarizes the information contained in the entire action sequence. We simplify the notation for the last recurrent state (i.e., the encoder output, $\boldsymbol\theta_{iL_i}$) to $\boldsymbol\theta_i$, which will be used to reconstruct the observed sequence in the decoding stage.

The first layer of the decoding stage is also a recurrent layer with $L_i$ time steps, with initial hidden state $\mathbf{y}_{i1} = \mathbf{0}$ and the $t$th hidden state obtained by $\mathbf{y}_{it} = f^\prime(\mathbf{y}_{it-1}, \boldsymbol\theta_i).$ Here, $f^\prime$ is another recurrent function, such as the GRU, with a different set of parameters from $f$. Note that the decoder recurrent states are obtained by feeding in the same encoder output, $\boldsymbol\theta_i$, for $L_i$ times and accumulating the information over time through $f^\prime(\cdot)$. 
The last layer of the decoder is a softmax layer, where, for each time point $1\leq t\leq L_i$, the decoder hidden state at time $t$ ($\mathbf{y}_{it}$) is used to predict the probability that individual $i$ takes each possible action. Intuitively, the decoder aims at reconstructing the distribution of the original sequence $\mathbf{s}_{i}$ using the encoder output $\boldsymbol{\theta}_i$.

\begin{figure}
    \centering
    \includegraphics[width = \textwidth]{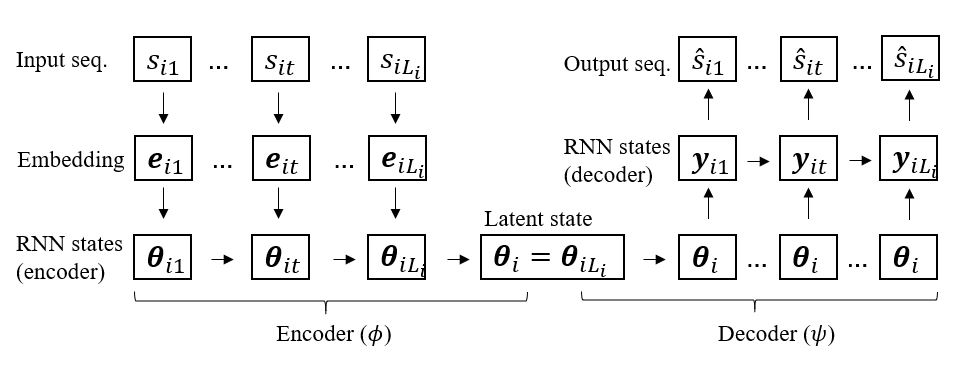}
    \caption{An illustration of the structure of the sequence-to-sequence autoencoder.}
    \label{fig:seq2seq_flow}
\end{figure}

A graphical illustration of the Seq2seq is presented in Figure \ref{fig:seq2seq_flow}. Because the output from the last time step of the encoder stage, $\boldsymbol\theta_i$, is used to reconstruct the original sequence in the decoder, $\boldsymbol\theta_i$ can be regarded as a summary of the entire sequence. For this reason, \citeauthor{tang2019seq2seq} \citeyear{tang2019seq2seq} proposed the use of $\boldsymbol\theta_i$ as the extracted features from the action sequences. For more information about the Seq2seq, including the explicit formula for each layer of the neural network and the details on model estimation, readers are directed to \citeauthor{tang2019seq2seq} \citeyear{tang2019seq2seq}.

As a concrete illustration of the extracted process features, Table \ref{tab: feature_example} presents the features extracted from three participants' processes on item U06b using either MDS or Seq2seq, with number of dimensions set as $K = 5$. Omitting ``Start'' and ``Next, Next\_OK'' at the beginning and the end, the three participants' action sequences were as follows:
\begin{itemize}
    \item Examinee 1: ``Click\_W4, Toolbar\_Web\_Back, Response\_Open, Response\_4, Response\_Close''; 
    \item Examinee 2:  ``Click\_W2'';
    \item Examinee 3: ``Click\_W1, Toolbar\_Web\_Back, Click\_W2''.
\end{itemize}
In practice, $K$ is often chosen to be larger than $5$ to preserve high-dimensional information in the action sequences. %\citeA{tang2019mds} proposed a method to choose $K$ through cross validations. 

\begin{table}[ht]
\caption{Example MDS and Seq2seq features extracted from U06b action sequences of $3$ examinees ($K = 5$).}
\label{tab: feature_example}
\centering
\begin{tabular}{rrrrrrr}
  \toprule
 & & MDS & & & Seq2seq &  \\\cmidrule(lr){2-4}\cmidrule(l){5-7}
 k & Examinee 1 & Examinee 2 & Examinee 3 & Examinee 1 & Examinee 2 & Examinee 3 \\ 
  \hline
1 & 0.04 & 0.16 & -0.04 & -0.56 & 0.09 & -0.62 \\ 
  2 & 0.15 & -0.27 & -0.15 & 0.90 & 0.08 & 0.89 \\ 
  3 & 0.20 & -0.07 & -0.04 & -1.00 & -0.90 & -1.00 \\ 
  4 & -0.07 & -0.09 & 0.13 & 0.94 & 0.62 & 0.94 \\ 
  5 & -0.01 & -0.17 & -0.21 & -0.72 & -0.36 & -0.73 \\ 
   \bottomrule
\end{tabular}
\flushleft
{\sl Note.} $k$: dimension of the MDS/Seq2seq features, ranging between $1$ and $5$.
\end{table}

\section{Method}

To investigate the associations between action sequences on each PSTRE item and various individual characteristics, features were first extracted from the action sequences of each item using MDS or Seq2seq. These features were subsequently used for the prediction of background variables and numeracy performance. For each predicted variable, partial least squares analysis was further performed on most predictive items' action features, so that the principle features that best explain the covariance between the predicted variable and the processes on the item were extracted and interpreted. In what follows, we provide a detailed description of the procedures applied in the current study for feature extraction from action sequences, individual characteristics prediction, and feature interpretations.

\subsection{Feature Extraction}
For each of the $14$ PSTRE items, $K= 100$ features were extracted from the action sequences of $N = 3645$ subjects using MDS and Seq2seq, respectively. Feature extraction with MDS was implemented in R with C++ integrations \cite{eddelbuettel2011rcpp}. Feature extraction with Seq2seq was implemented in Python with the Keras module \cite{chollet2015keras}. %, which estimates the parameters for the recurrent neural network of the autoencoder and generates model-based predictions of the individual features, $\boldsymbol{\theta}$.
Parameters of the neural net were optimized by minimizing the categorical cross-entropy loss between the observed sequences and predicted distributions of each action at each time step in the decoder, and the RMSProp \cite{tieleman2012lecture} optimizer was used. Following feature extraction, principal component analysis \cite<PCA;>{pearson1901liii} was applied to the $100-$dimensional features of each item to obtain linearly independent principal features ordered by amount of explained variance.

\subsection{Prediction and Evaluation}

The $100$ principal features of each PSTRE item were used to draw predictions about the examinees' age, gender, transformed hourly income, use of ICT tools at home, use of ICT tools at work, years of education, and numeracy performance.
For each item, three different types of predictors were considered, namely (1) subjects' polytomous scores on the item, (2) the $100$-dimensional principal features of each subject extracted from action sequences using the Seq2seq, and (3) the $100$-dimensional principal features from MDS. In addition to predictions with the features from each single item, cumulative predictive powers of final scores, MDS and Seq2seq were also evaluated with features combined across multiple items. Specifically, polytomous score, MDS and Seq2seq features were added one item at a time to examine how prediction accuracy grows as the information from more items were used. Generalized linear models (GLMs) were fitted through the R \texttt{glmnet} package \cite{friedman2009glmnet} to make the predictions, with a logit link function for the prediction of gender and an identity link function for the predicting the continuous variables. An $L_2$ penalty was applied in the GLMs to avoid overfitting. 
To reduce Monte Carlo error, ten replications were carried out for each prediction task. Specifically, in each replication, the data were randomly partitioned into $3$ subsets, a training set ($70\%$), a validation set ($10\%$), and a test set ($20\%$). The parameters of the GLM were estimated based on the training data, the optimal penalty term for $L_2$ regularization was chosen to minimize the loss function on the validation data, and the prediction accuracy of the GLM was evaluated on the test data. The evaluation criteria of prediction accuracy were as follows: For gender, we calculated the area under the curve (AUC) of the receiver operator characteristic (ROC) curve on the test sample. For the rest of the predicted variables, we calculated the out-of-sample Pearson correlation ($O.S.R$), which is the correlation between the predicted and observed values of a dependent variable in the test sample.

\subsection{Feature Interpretations}
For selected items whose process features demonstrated especially high associations with a specific variable, interpretations for the process-variable associations were further sought. To do this, partial least squares \cite<PLS;>{wold2002partial} decomposition was applied to extract the top $M$ orthogonal components from the action-sequence features that could best account for the covariance between the features ($\mathbf{X}$) and the dependent variable ($Y$). 
PLS is commonly used for obtaining lower-dimensional representations of high-dimensional features. Unlike PCA, where the top $M$ principal components capture the most variance in $\mathbf{X}$, in PLS, the top $M$ components capture the most of the covariance between $\mathbf{X}$ and $Y$. Thus, by seeking an $M$-dimensional PLS approximation to the $100$-dimensional action sequence features with respect to a dependent variable of interest, we obtain a set of $M$ linearly independent components, each explaining some of the associations between the MDS/Seq2seq features and the dependent variable. After PLS component extraction, individual's action sequences were sorted based on each of the $M$ PLS components. By inspecting how the observed action sequences change as the PLS component score increases, specific pattern(s) that explain the covariance between process features and the predicted variable can be pinpointed.

In the current study, the dimension for the PLS approximation, $M$, was chosen as follows:  We first ran PLS with $100$ components. For each $M^\prime \in \{1,\ldots, 100\}$, we calculated the root-mean-squared-error for the prediction (RMSEP) of $Y$ using the first $1,\ldots, M^\prime$ PLS components as linear regression predictors, i.e., $RMSEP_{M^\prime} = \sqrt{\frac{\sum_{i=1}^N (\hat y_i - y_i)^2}{N}}$. Denote the number of dimensions with the absolute lowest RMSEP by $M^\prime_{min}$ and the standard error of RMSEP across all $M^\prime$s by $SE(RMSEP)$, then the optimal number of dimensions, $M$, was chosen to be the smallest number of dimensions with RMSEP within one standard error of $RMSEP_{M^\prime_\text{min}}$, in other words, $M = \min\{M^*: RMSEP_{M^*} <  RMSEP_{M^\prime_\text{min}} + SE(RMSEP)\}.$

For each of the $M$ PLS components, we searched for patterns in the action sequences associated with the component. To do this, we first ranked all test-takers based on the PLS component, then, from lowest to highest, at an interval of $50$, we inspected the observed sequences of the test-takers. By looking at how the sequences changed as the component score increased, patterns in the action sequences associated with the PLS component were identified. We further verify the relationship between the visually spotted patterns and the PLS component by plotting how the pattern changes as a function of the PLS component score. More specifically, at each PLS component score value, we looked at $100$ individuals whose observed PLS component score was closest to this value and calculated the proportion demonstrating a specific pattern or the average frequency of a specific pattern among these $100$ individuals. This proportion/averaged frequency is then plotted against the PLS score.

\section{Results}

\subsection{Prediction Accuracy}

\begin{figure}
    \centering
    \includegraphics[width = \textwidth]{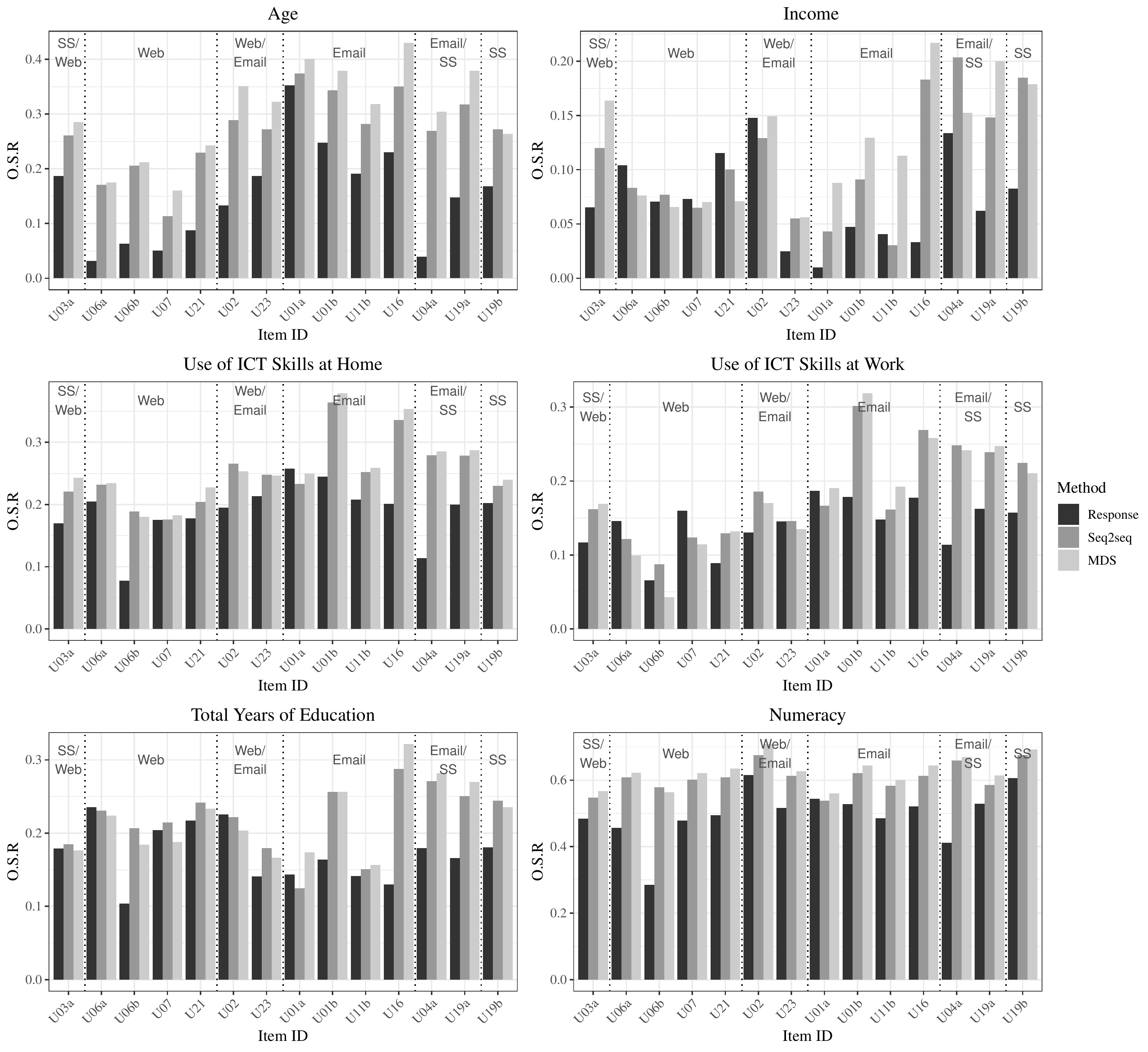}
    \caption{Prediction accuracy of continuous variables from polytomous scores, Seq2seq features, and MDS features of each PSTRE item. O.S.R stands for test-sample (out-of-sample) correlation between observed and predicted values.}
    \label{fig:single_pred_results}
\end{figure}

\begin{figure}
    \centering
    \includegraphics[width = \textwidth]{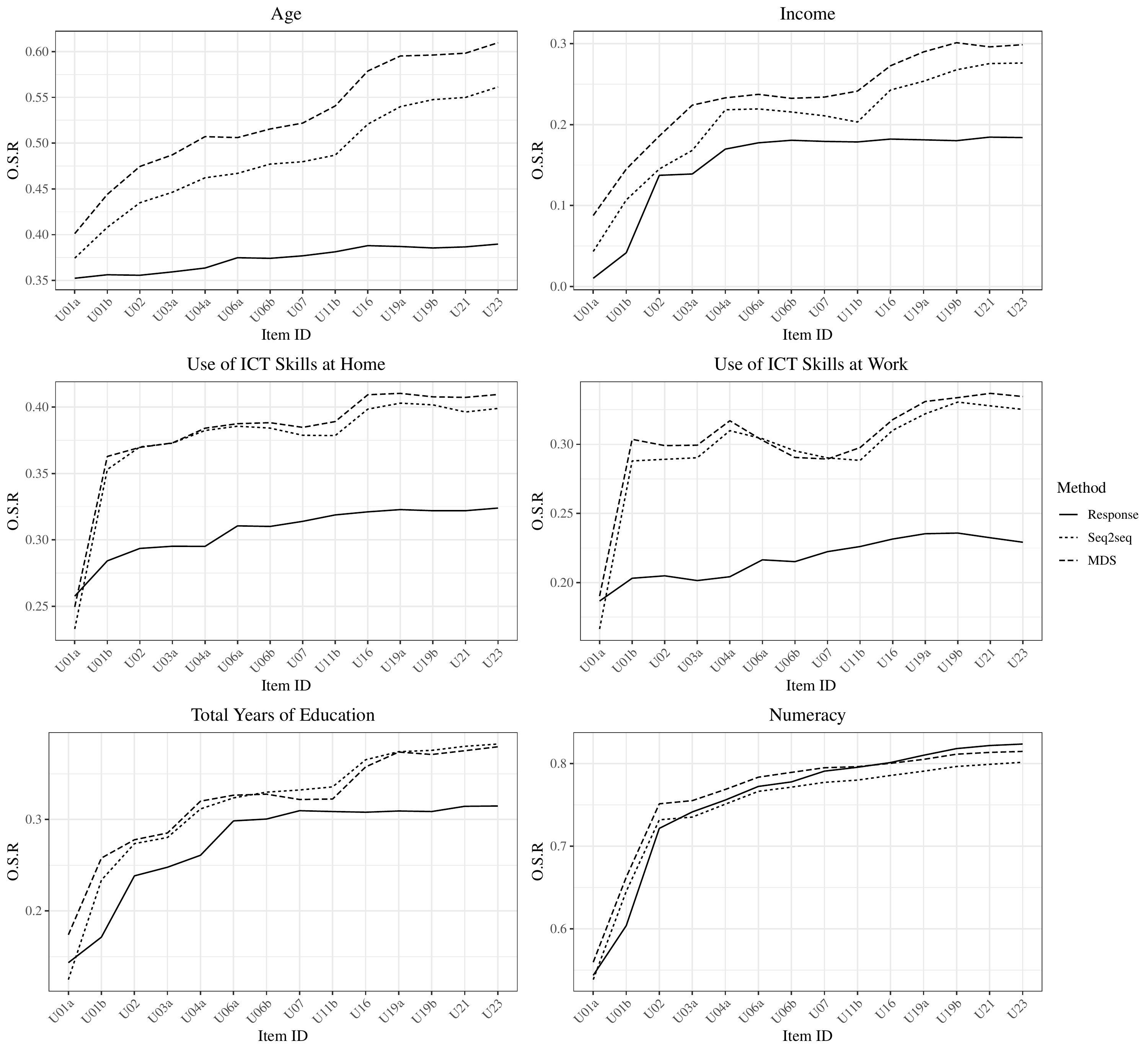}
    \caption{Cumulative prediction accuracy of continuous variables from polytomous scores, Seq2seq features, and MDS features with 1 to 14 items. Predictors from different items were incrementally added in the order on the x-axis. O.S.R stands for test-sample correlation between predicted and observed values.}
    \label{fig:cum_pred_results}
\end{figure}

\begin{figure}
    \centering
    \includegraphics[width = \textwidth]{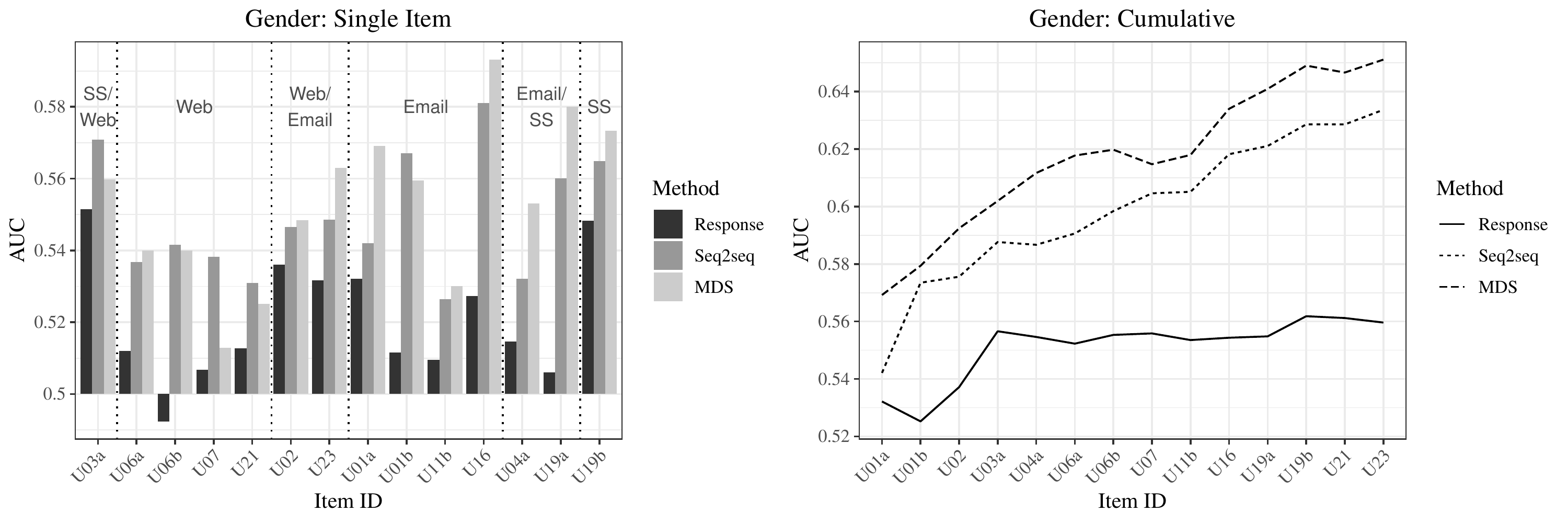}
    \caption{Single-item and multiple-item cumulative prediction accuracy of gender from polytomous scores, Seq2seq features, and MDS features.}
    \label{fig:gender_pred_results}
\end{figure}

The single-item prediction accuracy of the continuous variables and gender is reported in Figures \ref{fig:single_pred_results} and \ref{fig:gender_pred_results} (left), respectively. In each subplot, the $x$-axis represents the item used for prediction, the $y$-axis represent the evaluation metric, i.e., averaged test-sample $O.S.R$ or the AUC across $10$ replications, and the three bars with different shades represent the prediction results based on polytomous score, $100$ Seq2seq features, and $100$ MDS features, respectively. As was mentioned in the item descriptions, most of the PSTRE items involved one or two of the following environments: spreadsheet (SS), web browser (Web), or email box (Email), and items involving similar environments could share some common actions. To examine environment effect on the strength of associations, items are grouped by environments separated by vertical dashed lines. The combined prediction results with multiple items for the continuous variables and gender are presented in Figures \ref{fig:cum_pred_results} and \ref{fig:gender_pred_results} (right), respectively. Each line plot presents the increment of the cumulative prediction accuracy ($O.S.R$ or AUC) on a particular variable, as each item's features are additionally added to the predictors.

\subsubsection{Overall trends}
In most cases, for the prediction of a dependent variable with both single-item and multiple-item information, features derived from the action sequences (Seq2seq or MDS) achieved higher prediction accuracy compared to polytomous score(s) only. This suggests that the action sequences indeed contained additional information about different test-taker characteristics, which cannot be recovered solely based on the polytomous final scores. Compared to features extracted using Seq2seq, MDS features often demonstrated slightly higher prediction accuracy, with a few exceptions, e.g., the prediction of income using item U04a.

The prediction power of PSTRE items differed widely across background variables. For participants' age, the $O.S.R$ statistic of single item prediction could exceed $.41$. For the use of ICT skills at home and at work, as well as the total years of education of the subject, the highest $O.S.R$ statistics exceeded $.32$. The $O.S.R$ statistic for single item prediction of income was around $.21$ the highest. We observe a consistent but small improvement in the sensitivity of gender detection when process-based features were used. Note that when a simple random guess is taken, the AUC will be $.5$. With response only, the AUC of each item for gender prediction were close to $.5$, and by using the Seq2seq or MDS features, the AUC were up to $.59$. Generally, traits that were more correlated with overall PSTRE score (as shown in Table \ref{tab:bg_ques_descriptive}) were predicted better by the single items' features. However, although age, ICTHome, and YRSEdu all had correlations around $\pm .3$ with PSTRE score, the prediction power of single PSTRE items on age appeared to be higher than on the other two traits, suggesting that some specific item characteristics related to age might not be explained by overall PSTRE performance. 
%Note that prediction accuracy for numeracy performances were highest and that the cumulative prediction accuracy of polytomous scores for PSTRE performance was particularly high. This is expected, as the items were designated to measure PSTRE, and the PSTRE scores were directly derived from the polytomous scores.

\subsubsection{Differences across item environments} 
With $14$ items in total and only a few items per environment type, we were not able to conduct any statistical tests on the systematic differences in prediction power across different environment combinations. Nonetheless, we make a few remarks based on Figures \ref{fig:single_pred_results} to \ref{fig:gender_pred_results}, but by no means are the findings statistically significant. For the prediction of age, we observed that items involving an email environment tended to be more predictive than the others. For income, action sequences from items involving spreadsheets (spreadsheet only, spreadsheet and email, or spreadsheet and web browser) demonstrated stronger association with age than others, with the exception of item U16 (Email). As for ICT skill use at home and at work, action sequences from web browser-related items showed weaker association than email/spreadsheet-related items. Polytomous scores on web browser items predicted years of education better than other items, whereas the action sequence-derived features from items involving spreadsheets appeared more predictive of education than those from other items. We did not find any systematic differences across item environment types when it comes to the prediction of numeracy performance or gender.

\subsubsection{Heterogeneity across items}
Although all $14$ items were designed to measure the same trait (PSTRE), there was noticeable heterogeneity across items in the prediction of different variables. To start with, despite that features extracted from action sequences consistently outperformed polytomous scores in most prediction tasks, the magnitudes of the improvements differed vastly, even for some items sharing the same environments. For instance, items U01a and U16 both involved an email environment and achieved over $O.S.R > .38$ in age prediction using the MDS features from action sequences. However, $O.S.R$ of age prediction using U01a polytomous scores alone reached $.36$, while for $U16$, the age prediction based on polytomous score was only correlated with actual age at $.22$ on the test sample. Another example is the prediction of ICT skill use at home using two email items, U01a and U01b: The prediction power of polytomous scores on the two items were close (approximately $.22$), but with action sequence features, U01b showed much higher prediction power ($.37$) than U01a ($.24$). In other words, although the final responses of the two items were about equally associated with ICT use as home, the problem-solving behavior on U01b showed stronger association with ICT use at home. Second, for the prediction of a certain trait, items with high prediction power using process features not necessarily showed higher prediction power than the others using final responses. An example is item U04a for the prediction of overall numeracy performance: Using polytomous score only, this item was below average compared to the others in prediction accuracy, yet the Seq2seq and MDS features derived from the process data on this item achieved the one of the highest prediction accuracy among all items. This indicates that, when the purpose is to predict a particular trait, an item's problem-solving process could be highly informative even if its final response is not. Thirdly, when sequence-derived features were used for the prediction of a specific variable, the range of prediction power across the $14$ items could sometimes be quite large. Taking the single item prediction of overall numeracy performance as an example, the $O.S.R$ ranged from $.55$ (U06b) to $.7$ (U02).

\subsection{Interpretations for Associations}

From the results on the general prediction accuracy, it was observed that not only do sequence features provide additional information about the test-takers, the prediction powers of different items also differed widely. This calls for a closer look at action sequences that contribute to the additional predictive power. As the MDS features often demonstrated highest prediction power, for each continuous variable, we performed PLS decomposition on the MDS features from some of the most predictive items. The current subsection presents the findings from the PLS decomposition and interpretations of the PLS components. Results on age are presented in details, and for succinctness, results on the other variables are summarized. 

\subsubsection{Age} 
To understand the patterns in action sequences that were associated age, we performed PLS decomposition on $4$ items whose MDS features predicted age better than the others, namely U01a, U01b, U16, and U19a. Two PLS features were identified from U01a and U01b each, four PLS features were identified from U16, and three were identified from U19a. For each item and each identified PLS component, the original action sequences of the participants were sorted from lowest to highest based on the value on the PLS component, and visual inspection was performed on the sorted sequences to identify changes in action patterns as the PLS component score increased.

\begin{figure}
\centering
\begin{subfigure}[b]{.49\textwidth}
    \centering
    \includegraphics[width = \textwidth]{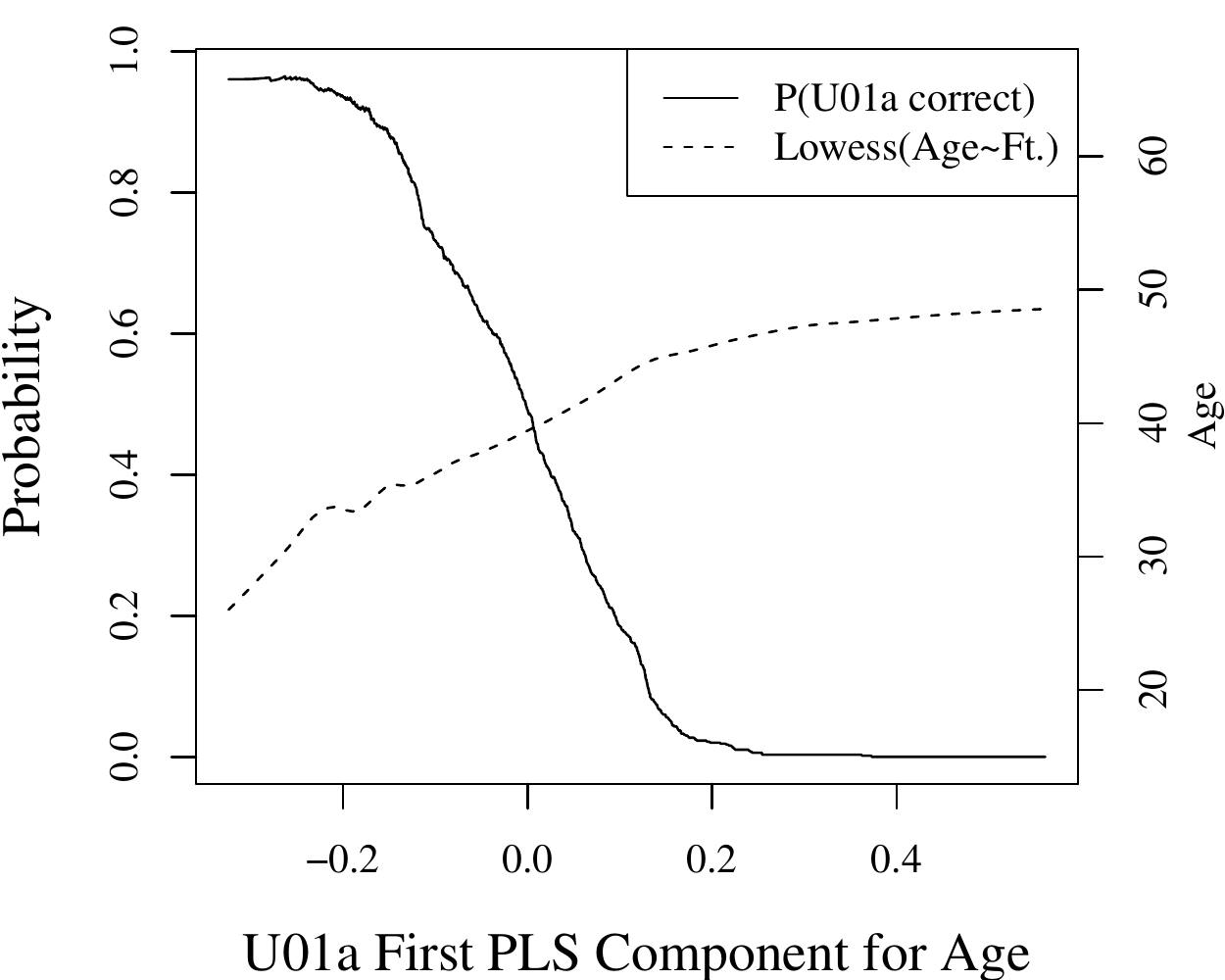}
\end{subfigure}
\begin{subfigure}[b]{.49\textwidth}
    \centering
    \includegraphics[width = \textwidth]{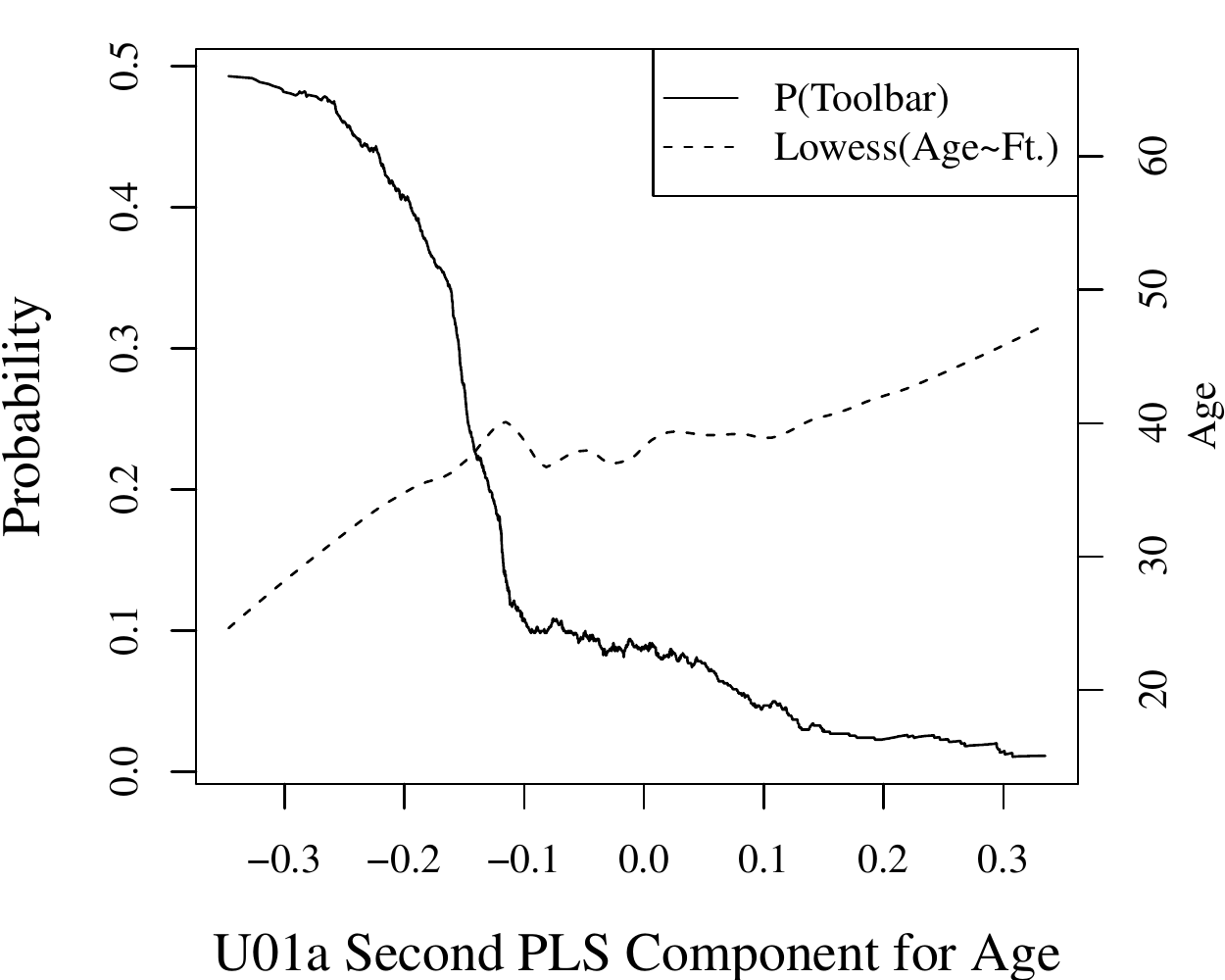}
\end{subfigure}
\caption{Distribution of U01a partial least squares component scores with respect to test-takers' age. Dashed lines on each subplot represent the local regression curve (LOWESS) of age against the PLS component score; solid lines on each curve presents how a pattern changes as PLS component score increases.}
\label{fig: PLS_U01a_age}
\end{figure}

\begin{figure}
        \begin{subfigure}[b]{.49\textwidth}
            \centering
            \includegraphics[width = \textwidth]{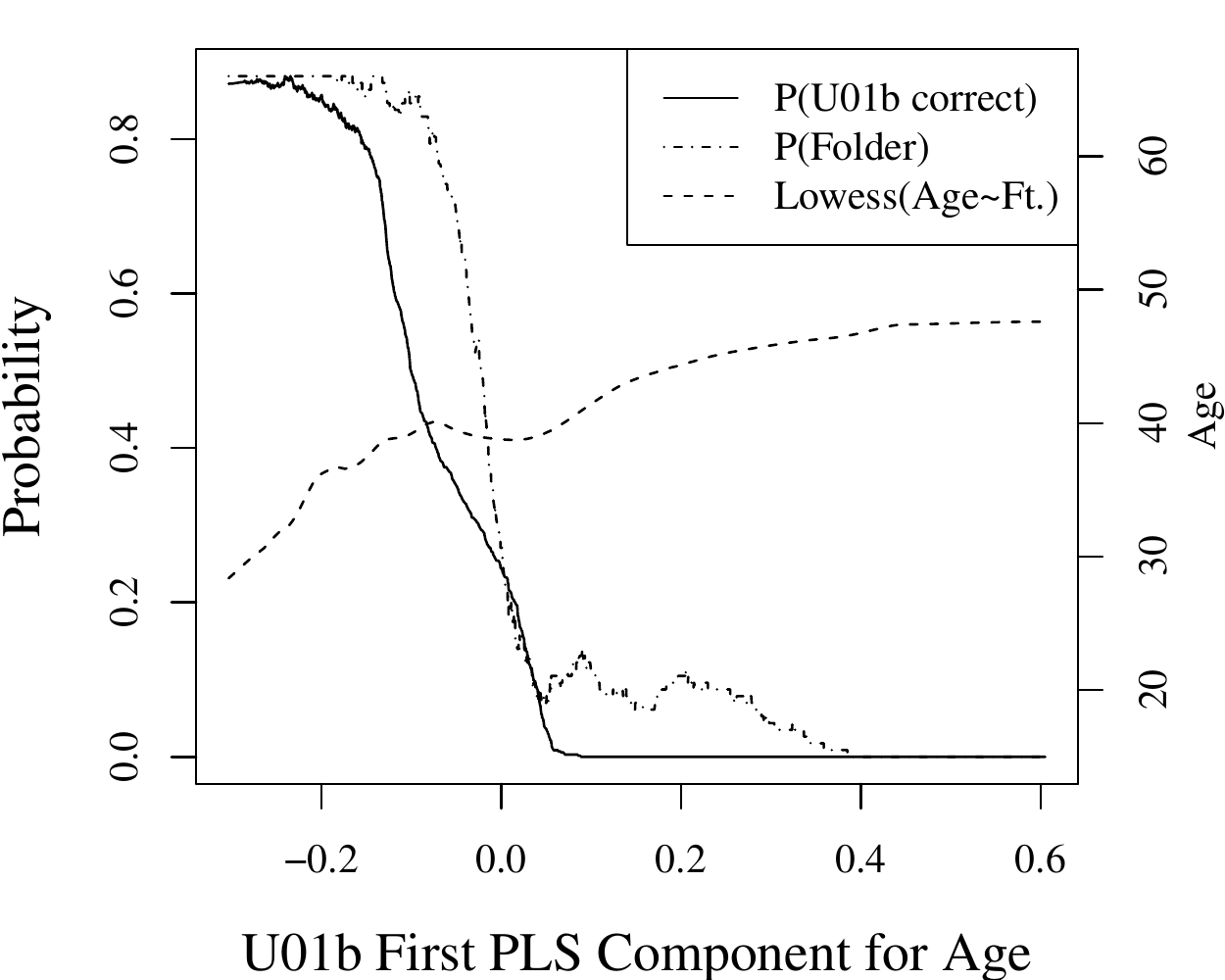}
        \end{subfigure}    
        \begin{subfigure}[b]{.49\textwidth}
            \centering
            \includegraphics[width = \textwidth]{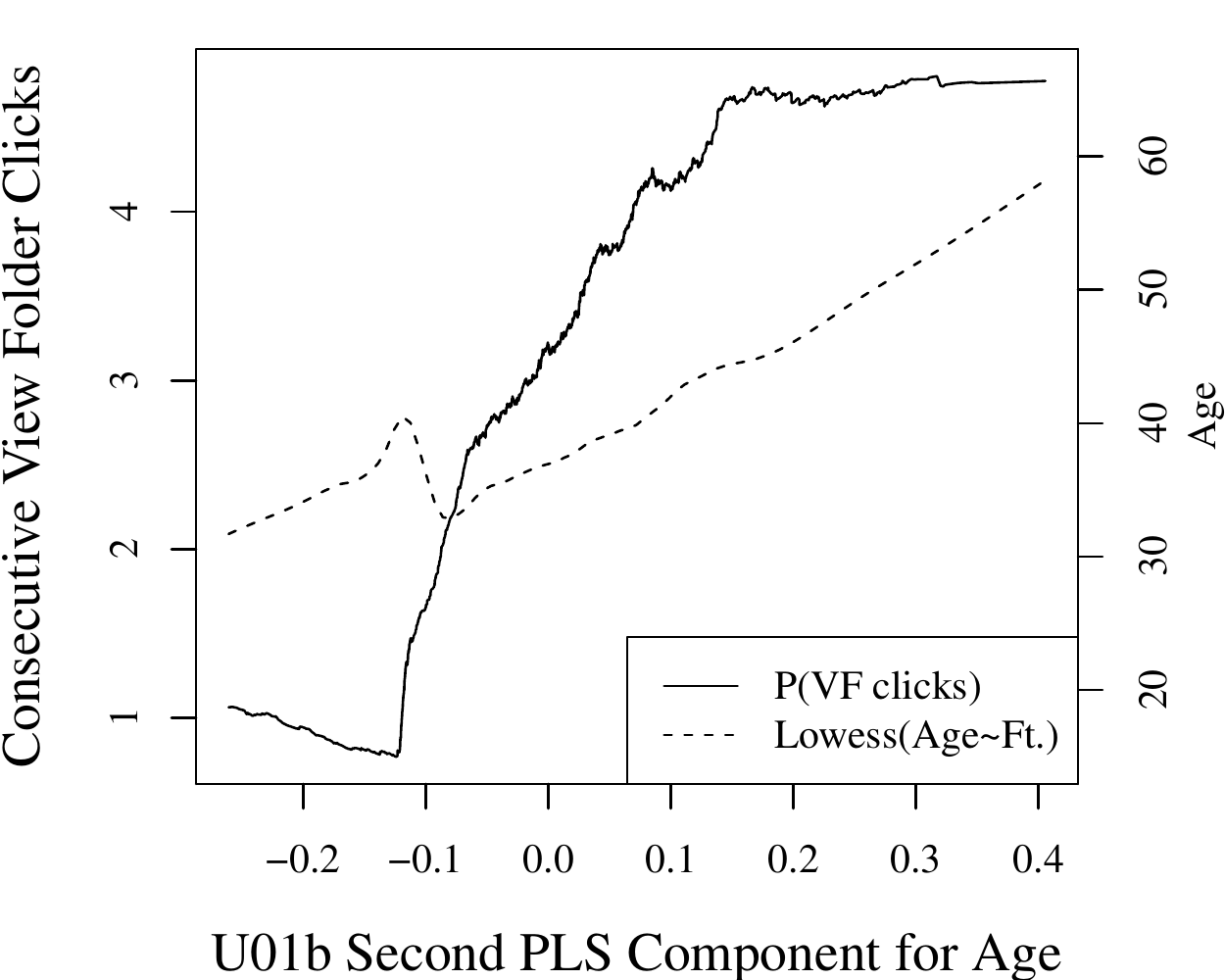}
        \end{subfigure}
        \caption{Distribution of U01b partial least squares component scores with respect to test-takers' age.}
        \label{fig: PLS_U01b_age}
\end{figure}

\begin{figure}
\begin{subfigure}[b]{.49\textwidth}
    \centering
    \includegraphics[width = \textwidth]{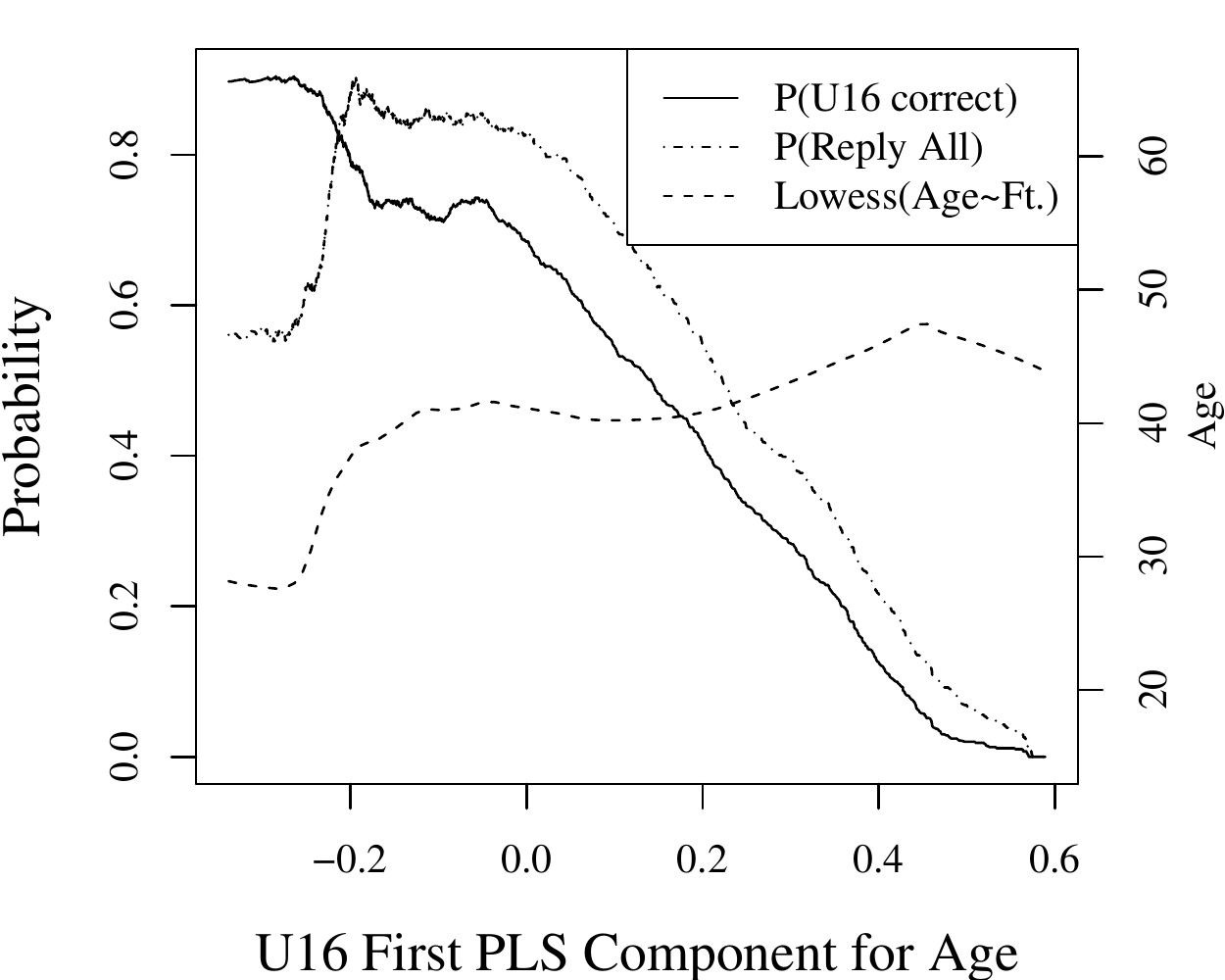}
\end{subfigure}   
\begin{subfigure}[b]{.49\textwidth}
    \centering
    \includegraphics[width = \textwidth]{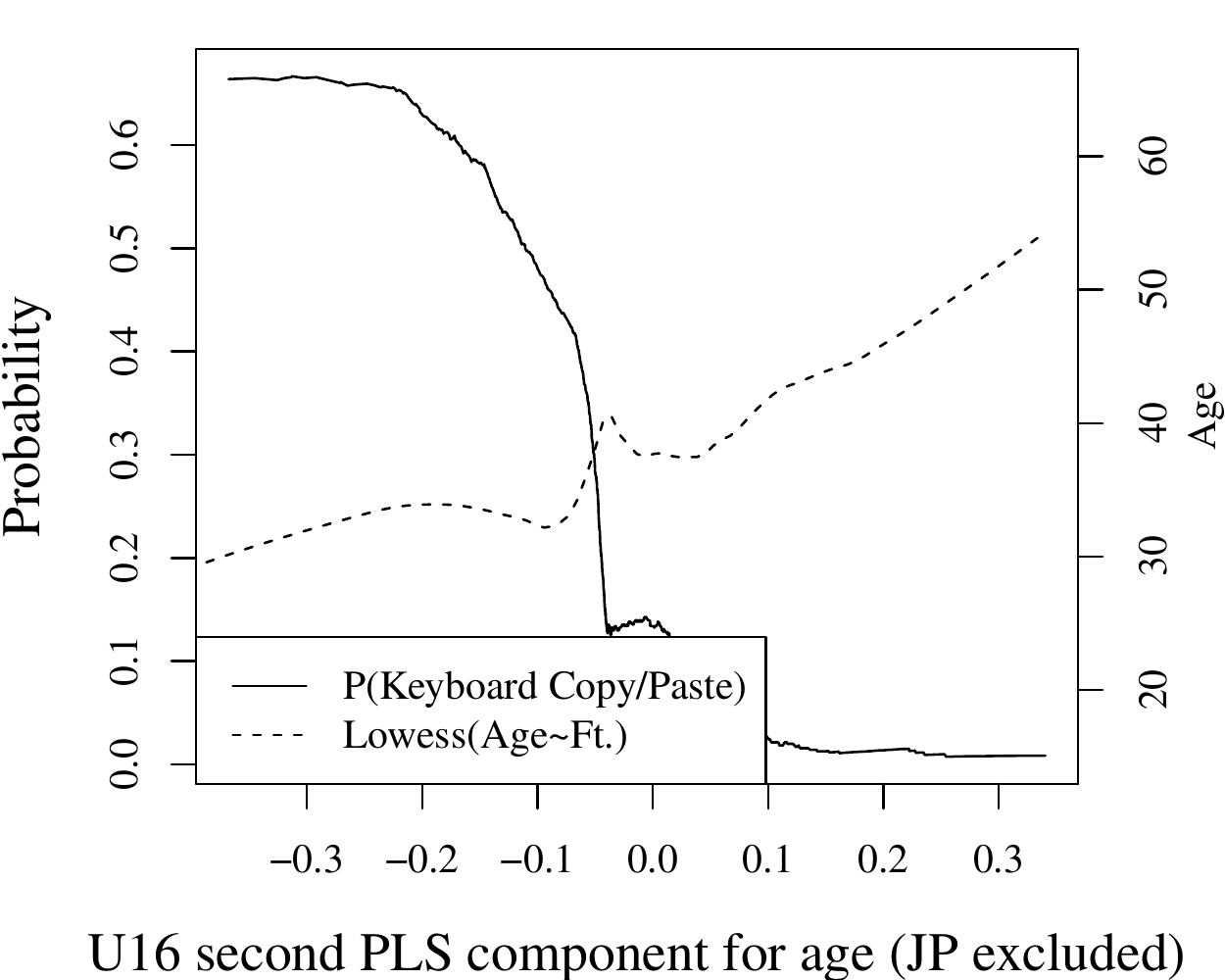}
\end{subfigure}
\begin{subfigure}[b]{.49\textwidth}
    \centering
    \includegraphics[width = \textwidth]{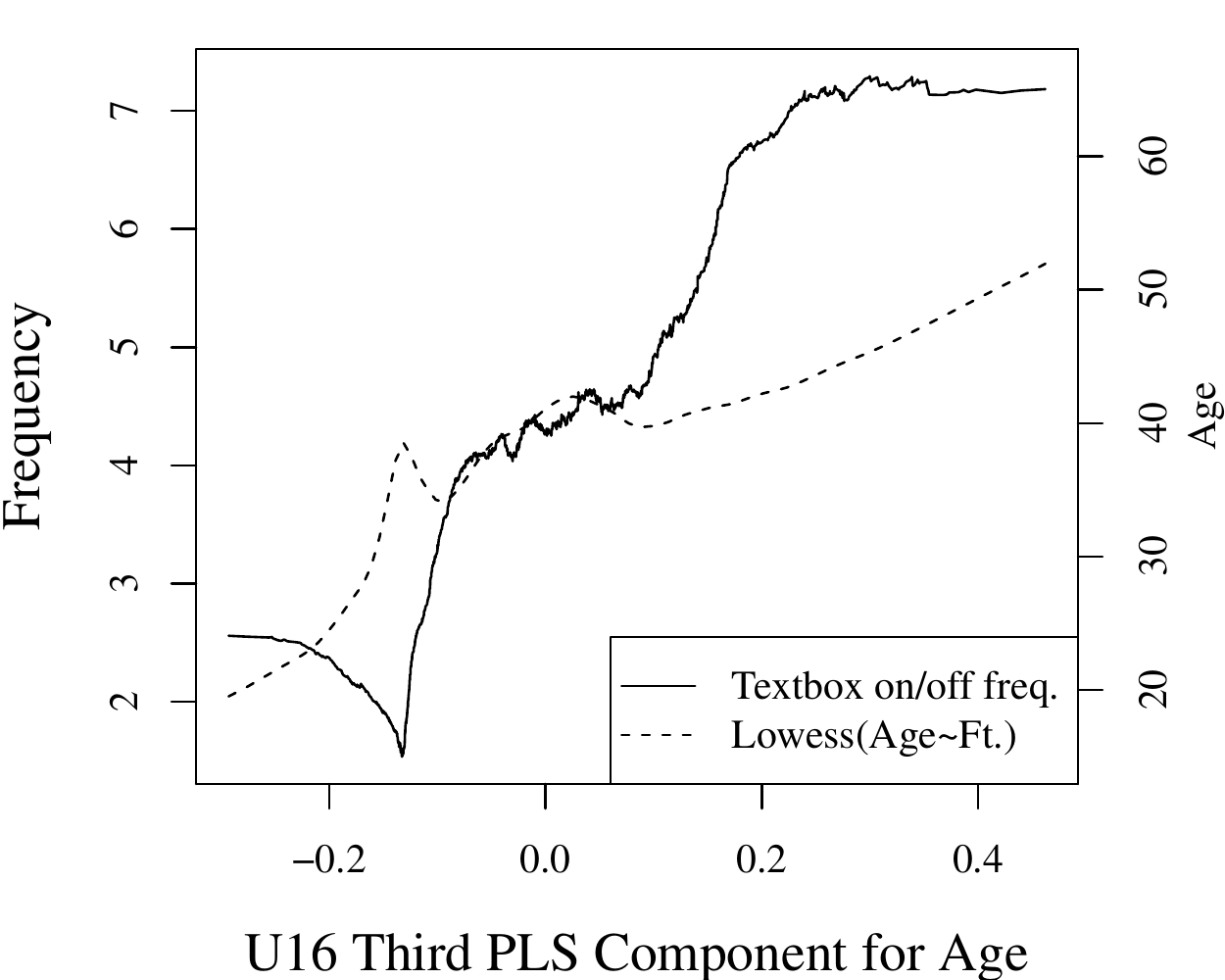}
\end{subfigure}    \begin{subfigure}[b]{.49\textwidth}
    \centering
    \includegraphics[width = \textwidth]{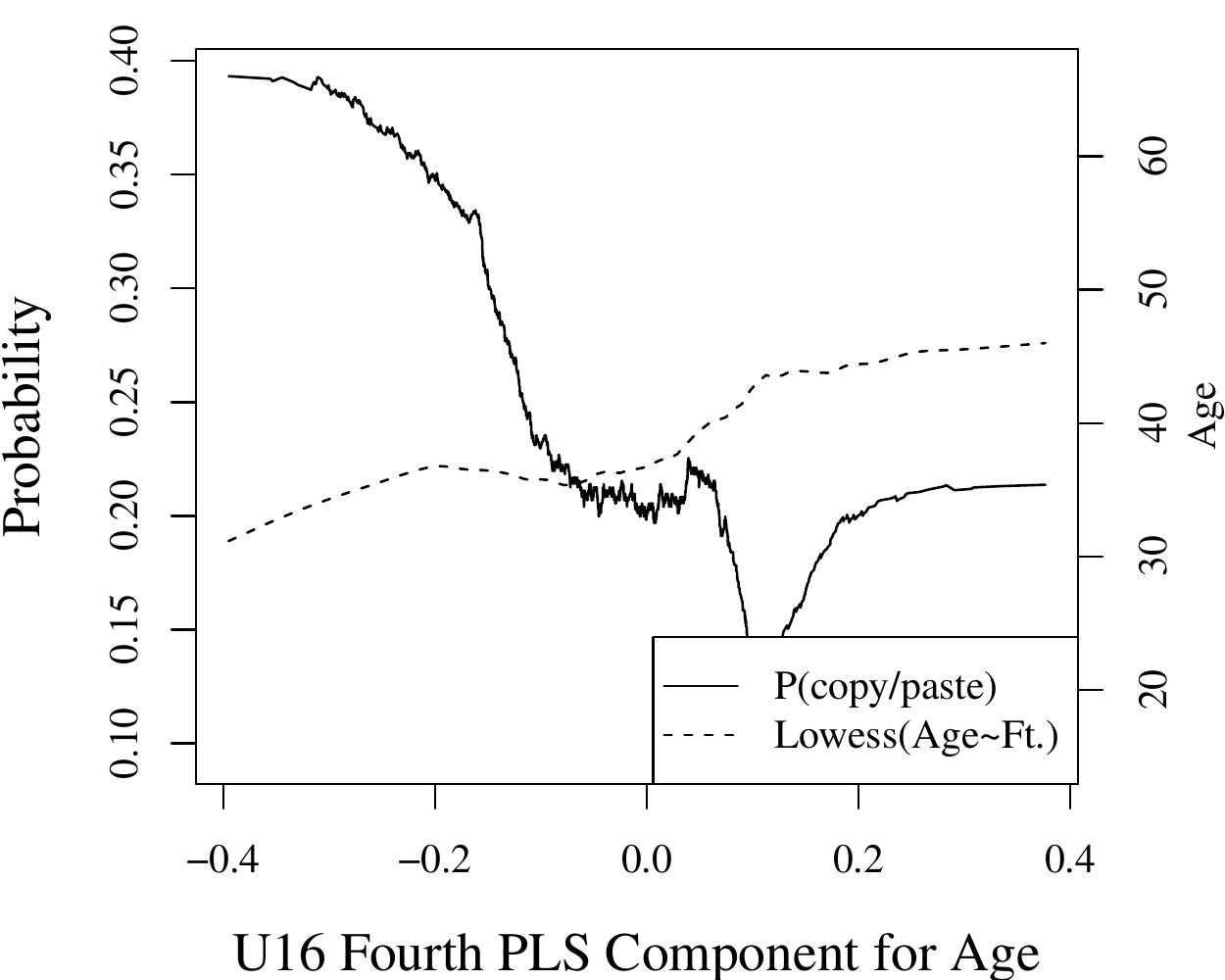}
\end{subfigure}
\caption{Distribution of U16 partial least squares component scores with respect to test-takers' age.}
\label{fig: PLS_U16_age}
\end{figure}

\begin{figure}\centering
\begin{subfigure}[b]{.49\textwidth}
    \centering
    \includegraphics[width = \textwidth]{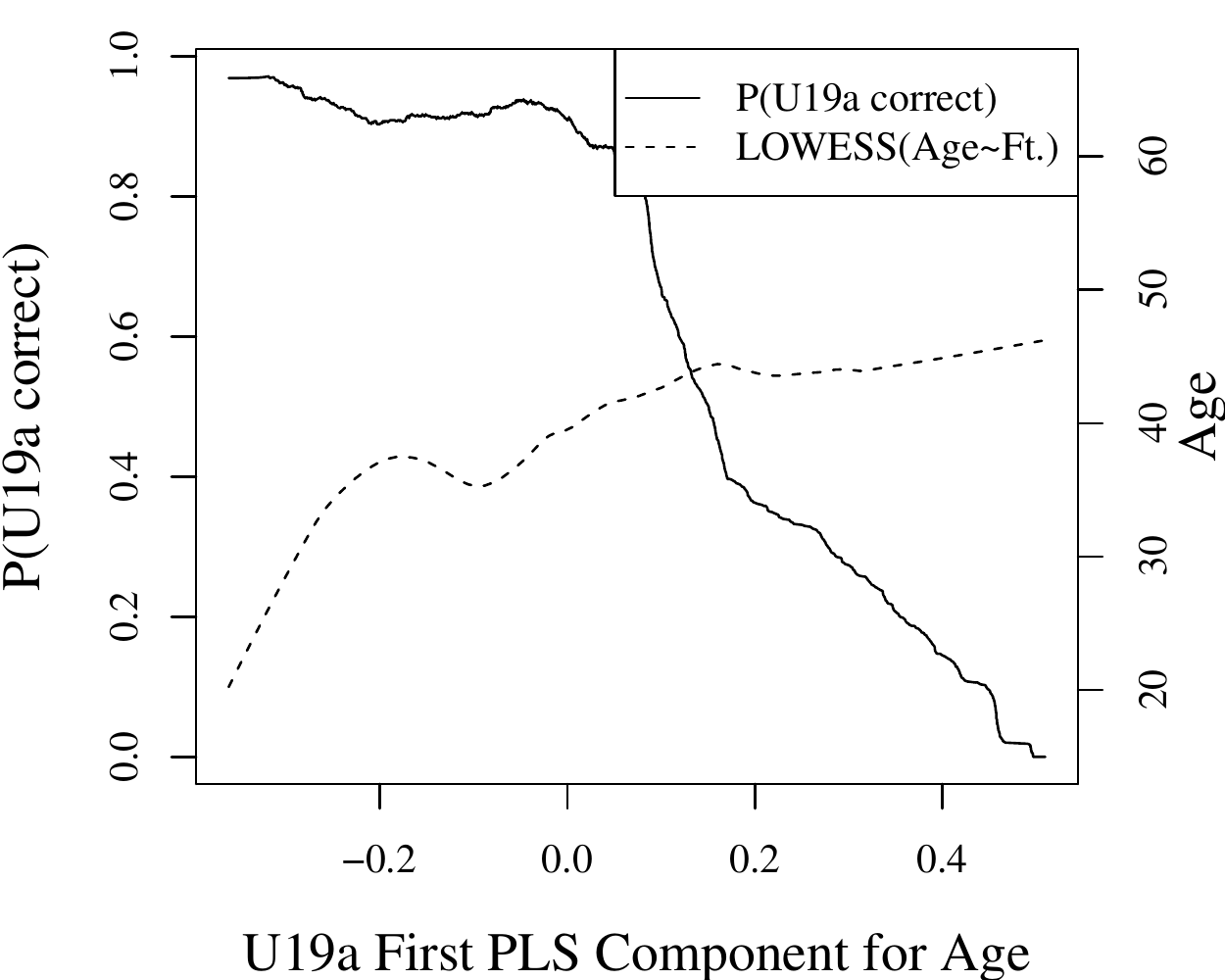}
\end{subfigure}    
\begin{subfigure}[b]{.49\textwidth}
    \centering
    \includegraphics[width = \textwidth]{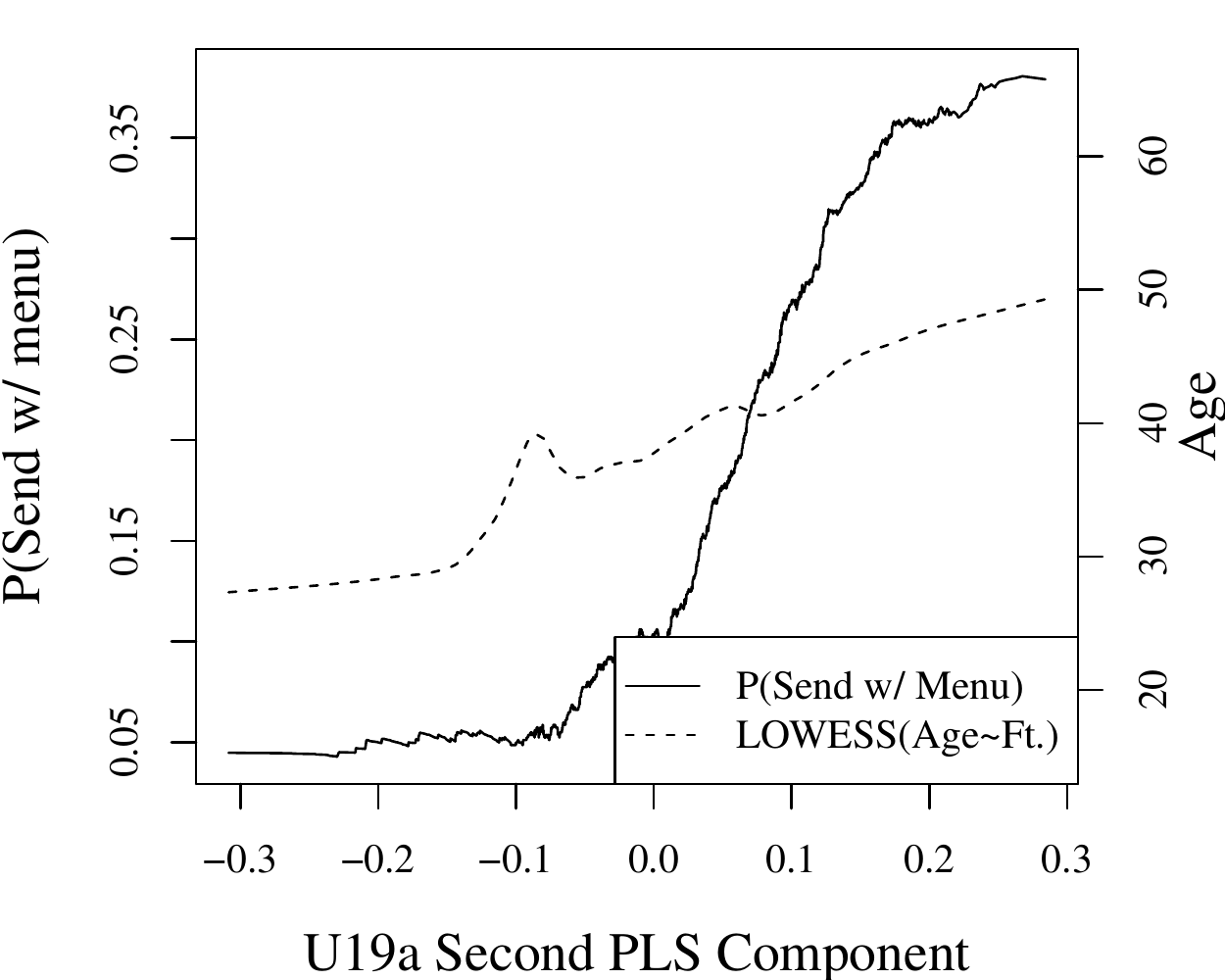}
\end{subfigure}
\begin{subfigure}[b]{.49\textwidth}
\centering
\includegraphics[width = \textwidth]{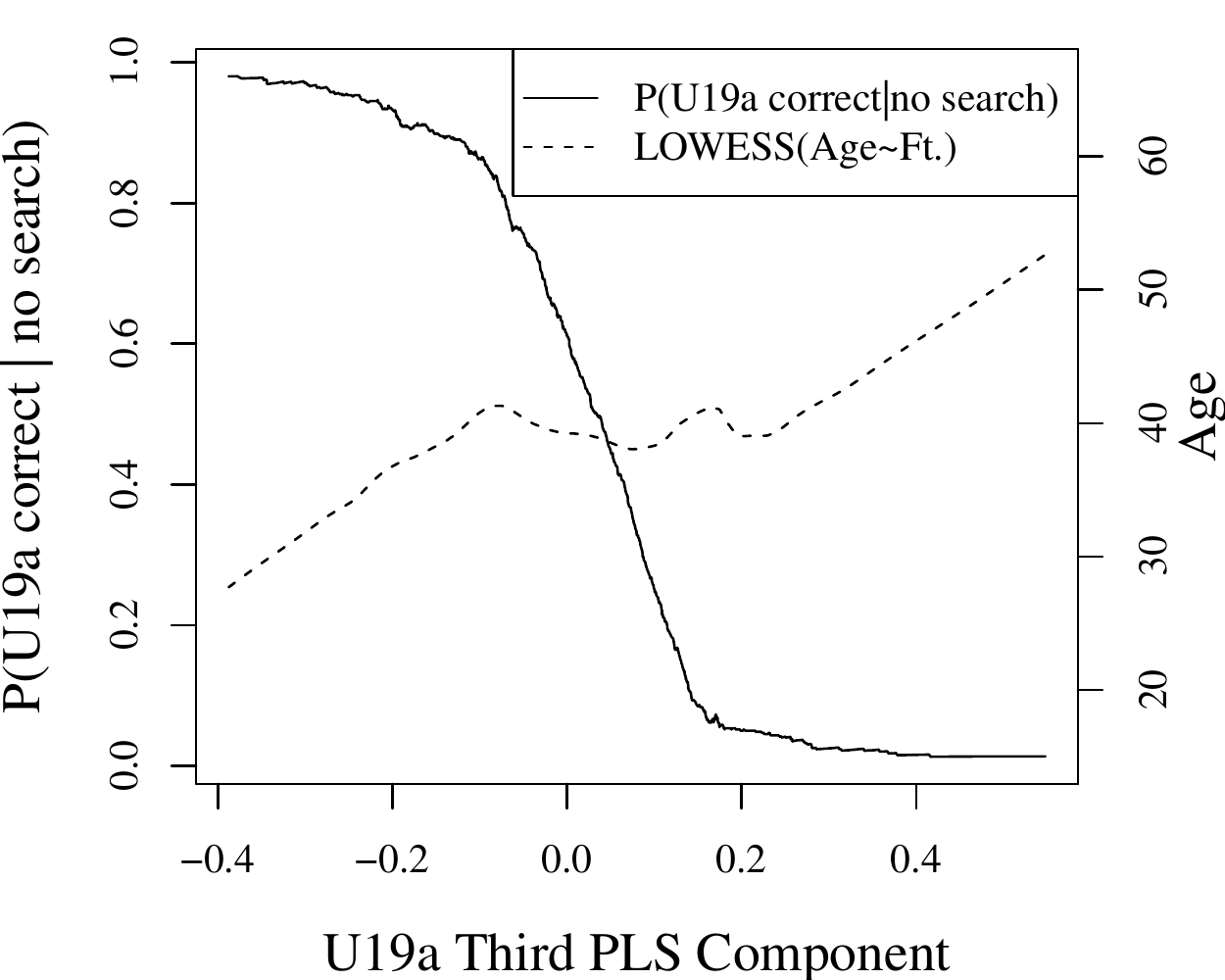}
\end{subfigure}
\caption{Distribution of U19a partial least squares component scores with respect to test-takers' age.}
\label{fig: PLS_U19a_age}
\end{figure}

Figures \ref{fig: PLS_U01a_age} to \ref{fig: PLS_U19a_age} present the interpretations found for each of the PLS components of the four items. Each subplot presents the relationship between the PLS component score with age, as well as the relationship between the component score with an observed action pattern, in the form of weighted local regression curves \cite<LOWESS;>{cleveland1981lowess}. The first PLS component from all four items, which explained the largest amount of covariance between MDS process features and age, were found negatively related to the probability of correct response. For items U01b and U16, the first PLS component was additionally found to be negatively related to creating a new email folder and using the ``Reply to all'' functionality in email clients, respectively. In other words, individuals higher on these components were less likely to respond correctly and were less likely to create a folder and use ``Reply to all''. The first PLS component on these items generally increased as age increased, suggesting that more senior individuals were generally higher on the component. 

Other PLS components can be interpreted in a similar manner. The second PLS component from U01a and the second component from U19a, both increased with age, were found to be negatively associated with the use of toolbar icons and positively associate with the use of drop-down menus, respectively. Note that for most PSTRE items, individuals could perform a task (e.g., sort a spreadsheet) either by clicking on an icon in the toolbar (e.g., clicking an arrow icon for sort) or by using the text-based drop-down menu. The second PLS components from U01a and U19a hence suggest an increasing tendency to use the drop-down menu rather than the toolbar icons as age increased. On item U16, which involved sending an email to a group of recipients, test-takers had the option to either hand-type the email or copy and paste the content from the question instructions. The second and fourth PLS components on U16 suggested a lower tendency to use copy/paste (either with keyboard shortcut or with menu/toolbar) in more senior individuals. Age was additional found to be positively related to consecutively clicking an email folder (U01b second PLS) and the number of times they switched back and forth between typing an email and looking at other contents (U16 third PLS). 

One special PLS component to note was the third PLS component on item U19a, where participants were prompted to identify a person's information from a spreadsheet. To identify the one person from a long spreadsheet, a test-taker could either scan through all rows, search for the person's name directly, or sort the spreadsheet alphabetically by names. The third component was found negatively related to the probability of responding correctly, given that the subject did not use the search bar. Test-takers on the low end of this component found the correct information of the individual without searching, while those on the high end very rarely responded correctly without using the search option. This component increased as age increased: Without searching for the individual directly, the test-takers would need to go through the long spreadsheet to find the individual, which may be more burdensome for senior test-takers.

\subsubsection{Income}
When it comes to the participants' hourly income, the MDS features from U16, U19a, and U19b demonstrated higher prediction accuracy compared to the others. The identified PLS components were uniformly found to be related to efficient strategies for problem-solving. Specifically, on U16, higher-income individuals were also higher on a PLS component that increased as the use of keyboard shortcuts (e.g., Ctrl+C, Ctrl+V) increased. On items U19a and U19b, the identified components (positively associated with age) were positively related the tendency to use search and sort in spreadsheets.

\subsubsection{ICT use at home and at work} 
PLS decomposition was performed on two items with relatively high prediction power on the use of ICT skills at home, namely U01b and U16. On item U01b, it was observed that individuals with more self-reported use of ICT tools at home were more likely to create email folders, which was a key step to correctly solving U01b. On U16, higher self-reported ICThome individuals were generally higher on PLS components for (1) more keyboard entry and (2) more use of keyboard shortcuts.

For self-reported use of ICT skills at work, we performed PLS analysis on one item achieving higher prediction accuracy than the others, U01b, and one PLS component was found, which was related to the probability of correct response to the item.

\subsubsection{Total years of formal education}

We report the PLS analysis results and interpretations of two items with relatively high prediction power on the test-takers' total years of education, namely U04a and U16. On U04, which involved creating a spreadsheet based on an email report, a PLS component positively related to years of education was found, which was associated with higher number of correct numerical entries in the spreadsheet. On U16, two PLS components positively associated with year of education were found, which were related to higher tendency of (1) using Cc and (2) using keyboard shortcuts for copy/paste, respectively. 

\subsubsection{Numeracy}
For numeracy, we focus on the interpretation of item U04, i.e., spreadsheet creation based on an email report. Three PLS components positively correlated with numeracy performance were found. The first PLS component was associated with the correct choice of spreadsheet row and column titles, with higher PLS score associated with higher number of correctly chosen spreadsheet row/column names. The second PLS component was associated with both correct table layout and correct numerical entries in the table: Test-takers on the low end specified only one factor correctly, that is, either correct on all row titles or on all column titles, but not both. And test-takers with high component scores were correct not only in row and column title specification but also in all numerical entries. The third PLS component appeared to be negatively related to the contrast between effort and performance, where higher PLS component score was associated with performing large number of actions (e.g., filling in spreadsheet titles and entries) without getting any spreadsheet entries correct.

\section{Discussions}

In this article, we examined to what extent action sequences from PIAAC PSTRE interactive items were associated with external background characteristics of the test-takers (i.e., age, gender, income, ICT use at home and at work, and years of education, and numeracy performance). Using the MDS and Seq2seq features extracted from the action sequences, we performed regression with $L_2$ regularization on various background and cognitive variables. One thing to note is that, although regression analyses were performed to draw predictions of external demographic and cognitive variables using process or response features, the purpose of such analyses was not to predict external variables with process data but to quantify the degree of association between problem-solving behavior and external covariates, in other words, the amount of information about an external trait that can be uncovered from the problem-solving processes.

A general tendency for the action sequence-derived features, especially the MDS features, to outperform polytomous responses alone on these predictions was observed, indicating that the sequences of actions an individual performs on an interactive item did indeed entail additional information on top of final responses. In addition, for a given dependent variable, the prediction power of the MDS/Seq2seq features differed vastly across items, and in some cases (e.g., the prediction of age), items involving specific environments (e.g., email) demonstrated higher overall prediction power than others. 

We further sought empirical interpretations when an item's MDS features achieved demonstrated higher association with a given trait compared to others. By performing partial least squares decomposition on the MDS features with respect to the external trait of interest, independent components that could explain the covariance between the MDS features and the trait were obtained, and we further sought patterns in the action sequences that corresponded to the components. While the action patterns we found to be associated with particular dependent variables differed across items, a few general observations could be made: (1) Whereas sometimes a background variable was found to be associated with the general performance on the item (e.g., response correctness on U01a, U01b, U16, and U19a for age), at other times, patterns that were associated with the background variable could, at least at first glance, seem unrelated to item or PSTRE performance (e.g., the choice of toolbar/drop-down menu use on U01a and U19a for age). (2) On items where multiple approaches to problem solving exists, whether an individual adopted a particular approach (e.g., reply-to-all in U16, keyboard shortcuts for copy/paste in U16, and searching/sorting in U19a/b) could reveal additional information about him or her (e.g., on income and ICT use at home and at work). (3) Even when an individual did not complete the key actions required for successful solution to the problems, the actions that they did perform could still be found associated with their cognitive performance (e.g., partially-correct table-filling on U04a).

With the rapid advancements in information technology and the evolving needs in educational assessment to measure new constructs (e.g., problem solving, technology literacy, and collaboration) through more complex and ecologically valid tasks, we start to see an increasing number of computer-based assessments with interactive items. While the measured constructs, form of delivery, structure of collected data, and methods for analysis and scoring may change, the fundamental characteristics that define assessment are unlikely to change \cite{bennett2018educational}: These characteristics include the careful design of items and tests to obtain evidence of examinee competency, the use of measurement models to connect observed data and individual/group characteristics of interest, and the evaluation of the quality of the assessments in terms of reliability and validity. Whereas traditional measurement models encounter barriers when it comes to the analysis of action sequences, methods such as MDS and Seq2seq can transform unstructured, variable length action sequence data into structured, fixed-dimensional features, which facilitates further predictions and inferences about the test-takers' cognitive performance and other characteristics. To developers of interactive assessments, the use of these action sequence-derived features goes far beyond predictions of examinee backgrounds or cognitive performance scores. More importantly, it offers a data-driven perspective on the evaluation, scoring, and design of interactive items and tests: First of all, by evaluating the prediction accuracy of a particular trait of interest based on extracted features from an item, we have a generic way to quantify the informativeness of the log data. Hence, if a subset of items are to be selected for the construction of a shorter test that can best differentiate examinees on a particular trait, priority may be given to items whose log data show higher prediction power. Secondly, when strong associations are found between specific patterns in the sequences and a measured construct, test developers may consider incorporating these patterns into the scoring of responses. In the design of future items, opportunities of collecting data on these highly informative patterns can also be planted, so as to increase the new items' expected informativeness. Thirdly, when associations are found between the features from an item and a background variable (such as age or gender), test developers can look into the specific action patterns that contribute to the associations, which can then be used to determine whether the differences in action sequences due to background characteristics is acceptable or should be construed as a peril to item fairness.

As an early attempt of using MDS/Seq2seq generated features to predict different test-taker characteristics and interpreting the associations between action sequences and background or cognitive variables, the current study has its limitations. To start with, for fair comparison between items, when we extracted features using MDS or Seq2seq from the action sequences, the dimension of the latent space was set uniformly to $K = 100$. As suggested in \citeauthor{tang2019seq2seq} \citeyear{tang2019seq2seq} and \citeauthor{tang2019mds} \citeyear{tang2019mds}, the optimal dimension $K$ can be chosen using cross-validations for better preservation of information in the sequences. Second, only linear models were considered in the prediction of different background and cognitive variables with the MDS and Seq2seq features. It is very possible that the relationship between certain features and the dependent variables is nonlinear, in which case the generalized linear model cannot adequately capture the associations. We leave the development and application of more complex models for the prediction of different test-taker characteristics from action sequence-derived features to future studies. Last but not least, pattern identification from the PLS components were performed by visually inspecting the ranked sequences. Not only is this process labor intensive, when the patterns are not as apparent, they could be missed by human eyes. Whereas methods such as n-gram analysis can be used to automatically identify subsequences of actions that are strongly associated with a component, feature, or variable \cite<e.g.,>{he2018exploring}, computational difficulties arise when longer subsequences are to be examined (e.g., when $n\geq 5$), making it hard for n-gram analyses to capture long-term information within the sequences. Furthermore, other than the presence or frequency of certain subsequences of actions, patterns in the action sequences can take other different forms, which cannot be pinpointed merely by the modelling of subsequences. As a direction for future research, methods for automatic pattern identification from features can be developed.

\bibliographystyle{apalike}
\bibliography{process_Ref}

\end{document}